\documentclass[aps,preprint,nofootinbib,superscriptaddress]{revtex4}

\usepackage{graphicx}
\usepackage{amsmath}
\usepackage{amsfonts}
\usepackage{bm}

\newcommand{\bea}{\begin{eqnarray}}
\newcommand{\eea}{\end{eqnarray}}
\newcommand{\beq}{\begin{equation}}
\newcommand{\eeq}{\end{equation}}
\newcommand{\bqa}{\begin{eqnarray}}
\newcommand{\eqa}{\end{eqnarray}}

\def\mqo2{{\!\!\!}}

\begin{document}

\title{The longitudinal response function of the deuteron in chiral
effective field theory}
\author{C.-J. Yang}
\email{cjyang@email.arizona.edu}
\affiliation{Department of Physics, University of Arizona, Tucson, AZ 85721, USA\\
}
\affiliation{Department of Physics and Astronomy and Institute of Nuclear and Particle
Physics, Ohio University, Athens, OH\ 45701, USA\\
}
\author{Daniel R. Phillips}
\email{phillips@phy.ohiou.edu}
\affiliation{Department of Physics and Astronomy and Institute of Nuclear and Particle
Physics, Ohio University, Athens, OH\ 45701, USA\\
}

\begin{abstract}
We use chiral effective field theory ($\chi$EFT) to make predictions for the
longitudinal electromagnetic response function of the deuteron, $f_L$, which
is measured in $d(e,e^{\prime}N)$ reactions. In this case the impulse
approximation gives the full $\chi$EFT result up to corrections that are of $\mathcal{O}(P^4)$
relative to leading order. By varying the cutoff in the $\chi$EFT
calculation between 0.6 and 1 GeV we conclude that the calculation is
accurate to better than 10\% for values of $\mathbf{q}^2$ within 4 fm$^{-2}$
of the quasi-free peak, up to final-state energies $E_{np}=60$ MeV. In these
regions $\chi$EFT is in reasonable agreement with predictions for $f_L$
obtained using the Bonn potential. We also find good agreement with existing
experimental data on $f_L$, albeit in a more restricted kinematic domain.
\end{abstract}

\maketitle

\smallskip 

\section{Introduction}

\label{sec:introduction}

The use of chiral perturbation theory ($\chi $PT) to describe few-nucleon
systems is a problem that has received much attention over the past twenty
years. This effort began with the seminal papers of Weinberg~\cite{We90},
and has recently been reviewed in Refs.~\cite{Ep08,Ep12}. In this approach
the NN potential is computed up to some fixed order, $n$, in the chiral
expansion in powers of $P\equiv (p,m_{\pi })/\Lambda _{0}$. Here $p$ is the
NN c.m. momentum, $m_{\pi}$ the pion mass, and the breakdown scale $\Lambda
_{0}$ is nominally $m_{\rho }\sim 4\pi f_{\pi }$, but in reality is somewhat
lower for reactions involving baryons. This NN potential is then iterated,
using the Schr\"{o}dinger or Lippmann-Schwinger equation, to obtain the
scattering amplitude. This approach has come to be known as chiral effective
field theory ($\chi $EFT) as it encodes the consequences of QCD's pattern of
chiral-symmetry breaking for few-nucleon systems and is built on a
systematic expansion in powers of $P$, while resumming the non-perturbative
effects that lead to the existence of nuclear bound states.

Electromagnetic reactions provide particularly fertile ground for the
application of $\chi $EFT. A few-nucleon electromagnetic current operator, $%
J_{\mu }$, can also be derived from $\chi $PT~\cite%
{Ko09,Ko11,Pa08,Pa09,Pa11,Pi12}, and matrix elements are then constructed
via: 
\begin{equation}
\mathcal{M}_{\mu }=\langle \psi ^{(f)}|\sum_{k=0}^{n}J_{\mu }^{(k)}|\psi
^{(i)}\rangle ,
\end{equation}%
where $|\psi ^{(i)}\rangle $ and $|\psi ^{(f)}\rangle $ are solutions of the
Schr\"{o}dinger equation for the $\chi $EFT potential $V$---in principle at
order $n$. $\chi $PT predicts that short-distance effects will enter $J_{\mu
}$ eventually, but minimal substitution generates the first several orders
in the expansion for $J_{\mu }$, especially in the case of the charge
operator. This enables predictions for charge form factors of $A=2$~\cite%
{PC99,WM01,Ph03,Ph07,Pi12} and $A=3$~\cite{Pi12} systems to be obtained as
pure predictions up to order $P^{4}$ relative to leading order. These
predictions agree well with data up to at least $Q^{2}=0.4$ GeV$^{2}$. In
the magnetic response the first short-distance operator not determined by NN
scattering or single-nucleon properties enters considerably earlier, but
accurate predictions for form factors can still be made~\cite{Ko12,Pi12}.
Indeed, the existence of this additional short-distance operator allows
enhanced understanding of magnetic moments and M1 transitions in systems up
to $A=9$, with reactions that had not previously been understood now being
explained, and correlated, through the presence of the same short-distance $%
\gamma $NN$\rightarrow $NN operator~\cite{Gi10,Pa12}.

In this work we focus on the application of this formalism to
electrodisintegration of deuterium, and specifically the computation of the
longitudinal response function, $f_{L}$. $f_{L}$ is proportional to the
square of the matrix element of the NN charge operator between the deuteron
wave function and the wave function of a continuum NN state. $J_{0}$ is
given solely by a single-nucleon operator up to $\mathcal{O}(P^{4})$
relative to leading order in the $\chi $EFT expansion, so calculations of $%
f_{L}$ can take the information gleaned on NN interactions from scattering
experiments and use it to make quite accurate predictions for an
electromagnetic observable. Deuteron electro-disintegration has been
employed as a testing ground for NN models for a long time~\cite%
{aren,aren1,aren2,ALT00,ALT02,ALT05}. Since the 4-momentum in the breakup
process is given by a virtual photon, a richer set of kinematics can be
probed as compared to the photo-disintegration process. And, in addition to
the deuteron wave function, the disintegration process probes both the
on-shell and off-shell NN t-matrix through the final-state interaction.

This means that the solutions for the deuteron wave function $|\psi \rangle$%
, and the NN t-matrix, are input to our calculation. To obtain them from $%
\chi$EFT potentials we need to impose a cutoff on the intermediate states, $%
\Lambda$. The low-energy constants (LECs) multiplying contact interactions
in the nucleon-nucleon part of the chiral Lagrangian should then be adjusted
to eliminate any cutoff dependence $\sim \mathcal{O}(1)$ or greater in the
effective theory's predictions for low-energy observables. If this is not
possible we conclude that $\chi$EFT is unable to give reliable predictions.
At leading order the $\chi$EFT potential consists of a zero-derivative
contact interaction that is operative only in NN partial waves with $L=0$
together with one-pion exchange, which is active in all partial waves. In an
earlier work we showed how to ``subtractively renormalize" the LO equations
for NN scattering in these two channels~\cite{Ya08}. This technique
eliminates the contact interaction in favor of a low-energy observable (e.g.
the relevant NN scattering length). This makes it straightforward to take
the limit $\Lambda \rightarrow \infty$: no fine-tuning of the pertinent LEC
is necessary. The resulting phase shifts (and one mixing parameter) do not
provide anything like a precision description of NN data, but they are, at
least, a well-defined, renormalized LO calculation.

However, several papers have demonstrated that a LO $\chi$EFT calculation
does not produce reliable predictions in partial waves with $L > 0$ once $%
\Lambda$ is sufficiently large~\cite{ES03,NTvK05,Bi06,PVRA06}. This is
because only one-pion exchange is present in these waves at LO, and the
resulting singular potential has no NN LEC that permits renormalization. We
examined this problem within the context of subtractive renormalization and
confirmed the conclusion of Ref.~\cite{NTvK05}, i.e. any partial wave with $%
L > 0$ where one-pion exchange is attractive does not have stable LO results
in $\chi$ET~\cite{Ya09A}. We also showed that this problem is not removed at 
$\mathcal{O}(P^2)$ or $\mathcal{O}(P^3)$. In particular, at $\mathcal{O}%
(P^3) $ two-pion exchange produces a highly singular, attractive potential.
The NN contact interactions needed to renormalize this potential are not
present, e.g. in the ${}^3$P$_2$-${}^3$F$_2$ channel. The resulting lack of
stability with $\Lambda$ of the NN phase shifts occurs once $\Lambda$ is
larger than $\approx 1$ GeV. In Ref.~\cite{Ya09B} we used subtractive
renormalization to calculate S-wave NN phase shifts from $\mathcal{O}(P^2)$
and $\mathcal{O}(P^3)$ $\chi$ET potentials. Here too we found that the phase
shifts are not stable once $\Lambda > 1$ GeV. In this case part of the
problem is that the momentum-dependent contact interaction that appears at $%
\mathcal{O}(P^2)$ has limited effect as $\Lambda \rightarrow \infty$~\cite%
{Wigner,PC96,Sc96}. Thus the NN potential obtained by straightforward
application of $\chi$PT cannot be used over a wide range of cutoffs: $\chi$%
EFT as formulated above is not properly renormalized, i.e. the impact of
short-distance physics on the results is not under control.

Nevertheless, in Refs.~\cite{Or96,Ep99,EM02} (Refs.~\cite{EM03,Ep05}) $V$
was computed to $\mathcal{O}(P^2)$ and $\mathcal{O}(P^3)$ ($\mathcal{O}(P^4)$%
), and the several NN LECs which appear in $V$ were fitted to NN data for a
range of cutoffs between 500 and 800 MeV. The $\mathcal{O}(P^4)$ predictions
contain very little residual cutoff dependence in this range of $\Lambda$'s,
and describe NN data with considerable accuracy. This suggests that $\chi$%
EFT may be a systematic theory of few-nucleon systems if we employ $\Lambda$%
's in the vicinity of $m_\rho$. Since the short-distance physics of the
effective theory for $p \gg m_\rho$ is different to the short-distance
physics of QCD itself, some authors argue that considering $\Lambda \gg
m_\rho$ does not yield any extra information about the real impact of
short-distance physics on observables~\cite{Le97,EM06}. Using low cutoffs
has the advantage that relevant momenta are demonstrably within the domain
of validity of $\chi$PT. Discussion about whether such a procedure results
in the omission of some operators is ongoing~\cite{Bi09,Ph12}. In what
follows we employ wave functions obtained with the standard $\chi$EFT
counting and our subtractive renormalization method~\cite{Ya09A,Ya09B}, but
we do so only for cutoffs $\Lambda$ up to the maximum value where we found
reasonable results in Refs.~\cite{Ya09A,Ya09B}, i.e. $\Lambda \le 1000$ MeV.

Our deuteron electrodisintegration calculation provides an opportunity to
test this strategy, by comparing its predictions for deuteron structure and
final-state interactions to experimental data. However, experimental data on 
$f_{L}$ are somewhat limited---especially in the low-energy and low-$\mathbf{%
q}^{2}$ region of most interest to $\chi $EFT. Much of the data on $f_{L}$
that do exist have been averaged over spectrometer acceptances, which makes
comparison with theory not only complicated, but also, in some cases,
ambiguous. Therefore, although we do compare with data from Refs.~\cite%
{vdS91,Du94,Jo96}, we use the Bonn-potential calculations of Arenh\"{o}vel 
\textit{et al.} as a proxy for data. These calculations have been quite
successful in describing data: non-L/T separated~\cite{Be81,TC84,Br86,Bl98},
L/T-separated~\cite{vdS91,Du94,Jo96}, and even on the interference response
functions, $f_{LT}$ and $f_{TT}$~\cite{Ta87,vdS92,Fr94,Ka97,Zh01,Bo08,Re08},
taken at many different laboratories over a wide range of relevant kinematic
conditions. (Note that we do not list data at squared momentum transfers $>1$
GeV$^{2}$ here because those are well outside the reach of $\chi $EFT, and
even beyond the scope of the semi-relativistic treatment of Arenh\"{o}vel 
\textit{et al.}. For recent progress on high-$Q^{2}$ electrodisintegration
see Refs.~\cite{Ul02,Bo11,JvO08,JvO10} and references therein.)

A necessary condition for the $\chi$EFT predictions to be considered
reliable is that they show minimal dependence on the cutoff $\Lambda$. We
will use this criterion to diagnose situations in which the final-state
interaction matrix-element computation has significant sensitivity to
short-distance physics. Deuteron photodisintegration has been studied
(albeit with an incomplete current operator) in $\chi$EFT by Rozdzepik 
\textit{et al.} using a similar strategy and the wave functions of Refs.~%
\cite{Ep05}~\cite{Ro11}. Since this process involves real photons, it is
sensitive to the deuteron transverse response function, $f_T$, and has no
dependence on the response function computed here. Christlmeier and Grie\ss %
hammer computed deuteron electrodisintegration at very low energies and
momentum transfers in the pionless effective field theory~\cite{CG08}. They
demonstrated the incompatibility of the data on the mixed response function $%
f_{LT}$, published by von Neumann-Cosel \textit{et al.} with the low-energy
NN phase-shift data and our knowledge of other electromagnetic transitions
in the NN system~\cite{vNC02}. This helped them diagnose a flaw in the
analysis of Ref.~\cite{vNC02}. Pionless EFT was also successfully applied to
the $(e,e^{\prime})$ data taken in a later experiment at S-DALINAC~\cite%
{Re08}.

The paper is structured as follows. In Sec.~\ref{sec-NNJ0OP3} we review the
facts about the NN charge operator which are relevant for our study. In Sec.~%
\ref{sec-d1}, we introduce the general set up of the electrodisintegration
problem and lay out the basic formulae. In Sec.~\ref{sec-d2} we evaluate the
matrix element of $J_0$ which enters $f_L$ and derive an explicit expression
for it in terms of partial waves. We also present the NN input used, i.e.
results for deuteron wave functions and NN phase shifts from Refs.~\cite%
{Ya09A,Ya09B}. In Section~\ref{sec-d4} we present our results for the
longitudinal response function. We pay particular attention to the kinematic
regions where significant dependence of $f_L$ on the momentum cutoff $%
\Lambda $ is and is not present. We present our conclusions in Sec.~\ref%
{sec-conc}.

\section{The NN charge operator to $\mathcal{O}(e P^3)$}

\label{sec-NNJ0OP3}

In this section we analyze the NN charge operator $J_{0}$, expanding it in
powers of the $\chi $EFT small parameter $P$. Since our wave functions are
computed up to $\mathcal{O}(P^{3})$ relative to leading order, we also need
the two-nucleon charge operator $J_{0}$ up to the same relative accuracy. We
denote the leading order for $J_{0}$ as $\mathcal{O}(e)$. A $J_{0}$ that is
consistent with chiral potential $V$ up to $\mathcal{O}(eP^{4})$ (and obeys
appropriate Ward identities) has recently been obtained by several authors~%
\cite{Ko09,Ko11,Pa09,Pa11,Pi12}. We now summarize the results needed for
this study.

The analysis proceeds by dividing the charge operator into an isoscalar part 
$J_{0}^{(s)}$, and an isovector one $J_{0}^{(v)}$. Up to corrections of $%
\mathcal{O}(eP^{4})$ in the chiral expansion we have, on the basis of states
of NN relative momentum~\footnote{%
We present $J_{0}$ in units of $|e|$, since these factors of the proton
charge are incorporated in the expression for the electrodisintegration
cross section.}:%
\begin{equation}
\langle \mathbf{p}^{\prime }|J_{0}^{(s)}(\mathbf{q})|\mathbf{p}\rangle =%
\left[ \delta (\mathbf{p}^{\prime }-\mathbf{p}-\mathbf{q/2})+\delta (\mathbf{%
p}^{\prime }-\mathbf{p}+\mathbf{q/2})\right] G_{E}^{(s)}(Q^{2})
\label{eq:J0s}
\end{equation}
where $G_{E}^{(s)}$ is the isoscalar nucleon form factor, and we have summed
over the contributions of nucleons one and two. In the case of the deuteron
we may use the symmetry of the state (only even partial waves are present)
to prove that the two terms in square brackets are equal.

Meanwhile for $J_{0}^{(v)}$ we have: 
\begin{equation}
\langle \mathbf{p}^{\prime }|J_{0}^{(v)}(\mathbf{q})|\mathbf{p}\rangle =%
\left[ \delta (\mathbf{p}^{\prime }-\mathbf{p}-\mathbf{q}/2)\tau
_{3}^{(1)}+\delta (\mathbf{p}^{\prime }-\mathbf{p}+\mathbf{q}/2)\tau
_{3}^{(2)}\right] G_{E}^{(v)}(Q^{2}),  \label{eq:J0v}
\end{equation}%
where $G_{E}^{(v)}$ is the isovector nucleon form factor. This operator does
not contribute to the matrix elements for elastic scattering, but it is
relevant for electrodisintegration. $G_{E}^{(s)}$ and $G_{E}^{(v)}$ are
related to the proton (neutron) charge form factor $G_{E}^{(p)}$ ($%
G_{E}^{(n)}$) by%
\begin{equation}
G_{E}^{(s,v)}(Q^{2})=\frac{1}{2}(G_{E}^{(p)}(Q^{2})\pm G_{E}^{(n)}(Q^{2})),
\label{form}
\end{equation}%
where the $+(-)$ sign applies to the $s(v)$ case.

$\chi $PT does a reasonable job describing $G_{E}^{(v)}$ for squared
4-momentum $Q^{2}<0.1$ GeV$^{2}$, but its description of isoscalar nucleon
structure $G_{E}^{(s)}$ is of limited utility, even in this low-$Q^{2}$
domain~\cite{Be92,Ph09}. Our goal here is to look at higher $|\mathbf{q}|$,
and we do not wish to be limited in our pursuit of that goal by $\chi $PT's
description of single-nucleon electromagnetic structure.

There are no two-body corrections to $J_0$ at $\mathcal{O}(eP^2)$. It might
seem that there will be an effect in $J_0^{(v)}$, because the $A_0$ photon
can couple directly to the exchanged pion. However, in the static limit used
to obtain NN potentials and charge operators that pion line carries no
energy. Thus two-body corrections to $J_0^{(v)}$ that contain such an effect
are deferred until $\mathcal{O}(eP^4)$, because we follow the counting of
Ref.~\cite{Ep05} and count $p/M \sim \mathcal{O}(P^2)$ while $p/\Lambda \sim 
\mathcal{O}( P)$. At $\mathcal{O}(eP^3)$ an NN contribution to $J_0$ must be
built out of a vertex from $\mathcal{L}_{\pi N}^{(2)}$, one from $\mathcal{L}%
_{\pi N}^{(1)}$ and a pion propagator. However, the only term in $\mathcal{L}%
_{\pi N}^{(2)}$ that couples an $A_0$ photon to a nucleon and a pion has a
fixed $1/M$ coefficient, so this effect is also deferred until $\mathcal{O}%
(eP^4)$.

Therefore, in the counting where $p/M \sim P^2$, the NN charge operator is
given by Eqs.~(\ref{eq:J0s}) and (\ref{eq:J0v}), up to corrections of $%
\mathcal{O}(eP^4)$.

\section{Basic formulae for deuteron electrodisintegration}

\label{sec-d1}

\begin{figure}[tbp]
\begin{center}
\includegraphics[width=14cm]{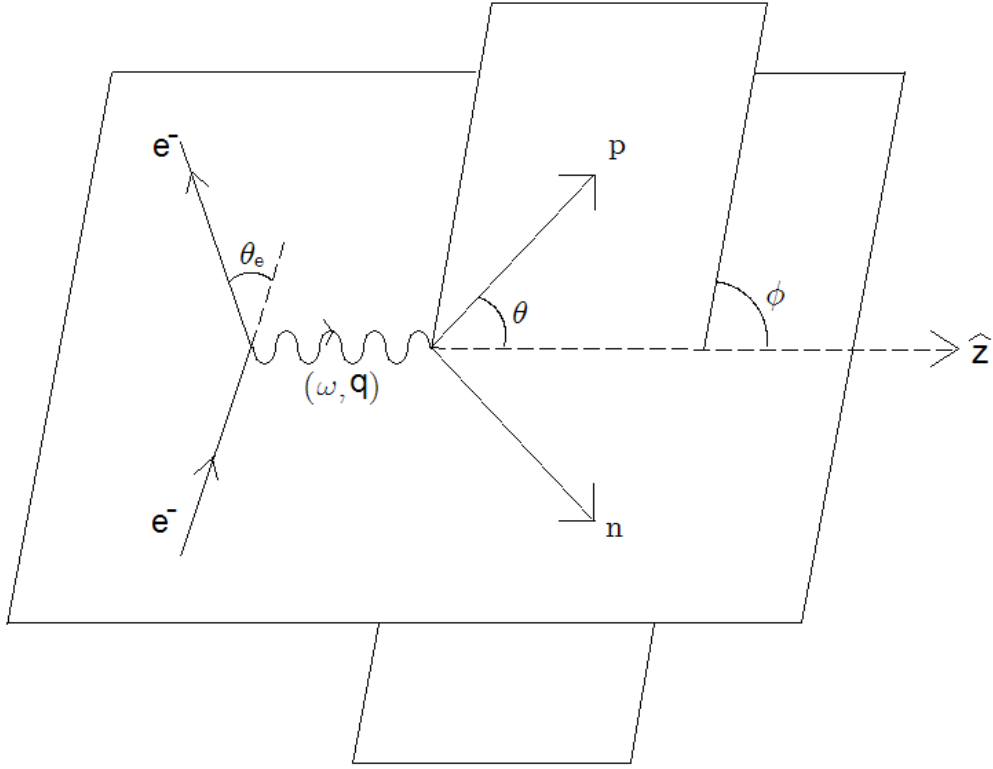}
\end{center}
\par
\vspace{3mm}
\caption{ The kinematic description of the deuteron electro-disintegration
process. The labels are specified in the text below.}
\label{fig-dkin}
\end{figure}

The usual expression for the differential cross section for deuteron
electro-disintegration is (see, for example, Ref. \cite{aren1})%
\begin{equation}
\frac{d^{3}\sigma }{dk_{2}^{lab}d\Omega _{e}^{lab}d\Omega _{p}}=c\{\rho
_{L}f_{L}+\rho _{T}f_{T}+\rho _{LT}f_{LT}\cos [\phi ]+\rho _{TT}f_{TT}\cos
[2\phi ]\}.  \label{eq:6.1}
\end{equation}%
$\rho (f)_{L,T,LT,TT}$ describe the lepton (hadron) tensor. The kinematics
is to be visualized as in Fig.~\ref{fig-dkin}. Here the virtual photon gives
its 4-momentum ($\omega ,\mathbf{q}$) to the deuteron, with these quantities
determined by the initial electron energy and $\theta _{e}$, the electron
scattering angle. $\theta $ is the angle between $\mathbf{q}$ and the
momentum $\mathbf{p}^{\prime }$ of the outgoing proton. (Here and in what
follows, unless otherwise stated, we work in the c.m. frame of the final
proton-neutron pair. If necessary, the superscript \textquotedblleft
lab\textquotedblright\ is used to denote quantities in the lab. frame.)
Meanwhile, $\phi $\ is the angle between the scattering plane containing the
two electron momentum vectors and the plane formed by the outgoing proton
and neutron. Finally, in Fig.~\ref{fig-dkin} we have defined $\mathbf{q}$ to
be along the z-axis. Hence $\Omega _{e}^{lab}=(\theta _{e}^{lab},\phi
_{e}^{lab}),$ $\Omega _{p}=(\theta ,\phi )$. Meanwhile, 
\begin{equation}
c=\frac{\alpha }{6\pi ^{2}}\frac{k_{2}^{lab}}{k_{1}^{lab}(q^{2})^{2}},
\end{equation}%
where $\alpha $ is the fine structure constant, $k_{1(2)}^{lab}$ is the
absolute value of the incoming (outgoing) electron 3-momentum in the
laboratory frame, and $q^{2}$ is the 4-momentum-squared of the virtual
photon.

In the (final-state) c.m. frame we have 
\begin{eqnarray}
\text{proton (neutron) 3-momentum} &:&\mathbf{p}^{\prime }\mathbf{(-p}%
^{\prime }\mathbf{),}  \notag \\
\text{proton (neutron) total energy} &:&E_{n}=E_{p}=M+\frac{\mathbf{p}%
^{\prime 2}}{2M},  \notag \\
\text{total energy of deuteron} &:&E_{d}=M_{d}+\frac{\mathbf{q}^{2}}{2M_{d}},%
\text{ }M_{d}=2M-B.  \label{eq:6.0}
\end{eqnarray}%
Here $M(M_{d})$ represents the rest mass of the nucleon (deuteron) and $%
B=2.24$ MeV is the deuteron binding energy. From energy conservation: $|%
\mathbf{p}^{\prime }|=\sqrt{(\omega -B)M+(\mathbf{q})^{2}\frac{M}{2M_{d}}}$.
The quantities in the lab. frame can be easily related to those in c.m.
frame by a Lorentz boost \cite{CG08} by an amount 
\begin{equation}
\beta =\frac{|\mathbf{q|}^{lab}}{M_{d}+\omega ^{lab}},  \label{eq:beta}
\end{equation}
i.e., 
\begin{eqnarray}
\omega^{cm} &=&\gamma \omega ^{lab}-\beta \gamma |\mathbf{q|}^{lab},  \notag
\\
|\mathbf{q|}^{cm} &=&\beta \gamma M_{d}.  \label{eq:LT}
\end{eqnarray}

For this work, we do not consider the electron polarization degree of
freedom, and calculate only the contribution from the longitudinal part in
Eq.~(\ref{eq:6.1}), which is thus rewritten as%
\begin{equation}
\frac{d^{3}\sigma _{longitudinal}}{dk_{2}^{lab}d\Omega _{e}^{lab}d\Omega _{p}%
}=\frac{\alpha }{2\pi ^{2}}\frac{k_{2}^{lab}}{k_{1}^{lab} (q^2)^2}%
\sum\limits_{SM_{s}m_{J}}\rho _{L}T_{SM_{s}0m_{J}}T_{SM_{s}0m_{J}}^{\ast },
\label{eq:6.5}
\end{equation}%
where the lepton tensor%
\begin{equation}
\rho _{L}=4E_{e}^{lab}E_{e}^{\prime lab}(\frac{q^{2}}{\mathbf{q}^{2}})\cos
^{2}\left( \frac{\theta _{e}^{lab}}{2}\right) ,
\end{equation}%
and $T_{SM_{s}\mu m_{J}}$ is defined as~\cite{aren}:%
\begin{equation}
T_{SM_{s}\mu m_{ds}}=-\pi \sqrt{2\alpha |\mathbf{p}^{\prime
}|E_{p}E_{d}/M_{d}}\langle \Psi _{\mathbf{p}^{\prime }SM_{s}}|J_{\mu }(%
\mathbf{q})|m_{J}\rangle .  \label{eq:6.4}
\end{equation}

In Eq.~(\ref{eq:6.4}) $\langle \Psi _{\mathbf{p}^{\prime }SM_{s}}|$ is the
NN final state with the total spin quantum number $S$ and its projection on
the z-axis, $M_{s}$, both specified, $\mathbf{p}^{\prime }$ represents that
the final proton has 3-momentum $\mathbf{p}^{\prime }$, and $\mu $ labels
the polarization index of the virtual photon. We consider only $\mu =0$
here. The deuteron state $|m_{J} \rangle$ has total angular momentum 1, and $%
m_{J}$ labels the z-projection of its total angular momentum. The angular
dependence of Eq.~(\ref{eq:6.4}) can be separated into two parts, i.e., 
\begin{equation}
T_{SM_{s}\mu m_{J}}(\theta ,\phi )=e^{i(\mu +m_{J})\phi }x_{SM_{s}\mu
m_{J}}(\theta ).
\end{equation}%
The longitudinal structure function $f_{L}$ is obtained from the $\theta $%
-dependent part of $T_{SM_{s}0m_{J}}$, i.e.,%
\begin{equation}
f_{L}=\sum\limits_{SM_{s}m_{J}}x_{SM_{s}0m_{J}}x_{SM_{s}0m_{J}}^{\ast },
\label{eq:6.5a}
\end{equation}
and so $\frac{d^{3}\sigma _{longitudinal}}{dk_{2}^{lab}d\Omega
_{e}^{lab}d\Omega _{p}} \sim f_L$, with the proportionality determined
solely by kinematic factors.

\section{Evaluating the matrix element}

\label{sec-d2}

From Eqs.~(\ref{eq:6.4}) and (\ref{eq:6.5a}), one sees that to obtain $%
f_{L}, $ the matrix element $\langle \Psi _{\mathbf{p}^{\prime
}SM_{s}}|J_{0}(\mathbf{q})|m_{J}0\rangle $ needs to be evaluated, which, up
to $\mathcal{O}(eP^{4})$, can be represented by 
\begin{eqnarray}
\langle \Psi _{\mathbf{p}^{\prime }SM_{s}T}|J_{0}(\mathbf{q})|m_{J}0\rangle
&=&\langle \mathbf{p}^{\prime }SM_{s} T|J_{0}(\mathbf{q})|m_{J}0\rangle 
\notag \\
&&+\langle \mathbf{p}^{\prime }SM_{s} T|t(E^{\prime })G_{0}(E^{\prime
})J_{0}(\mathbf{q})|m_{J}0\rangle ,  \label{eq:6.6a}
\end{eqnarray}%
with $J_0$ given by Eqs.~(\ref{eq:J0s}) and (\ref{eq:J0v}), and $t(E^{\prime
})$ and $G_{0}(E^{\prime })$ the NN t-matrix and free Green's function. Here
we have used the $T$ to represent the fact that the final-state wavefunction
is isospin dependent and introduced a $0$ in the kets and the $T$ in the two
bras on the right-hand side to indicate the isospin of those states.

\begin{figure}[tbp]
\begin{center}
\includegraphics[width=14cm]{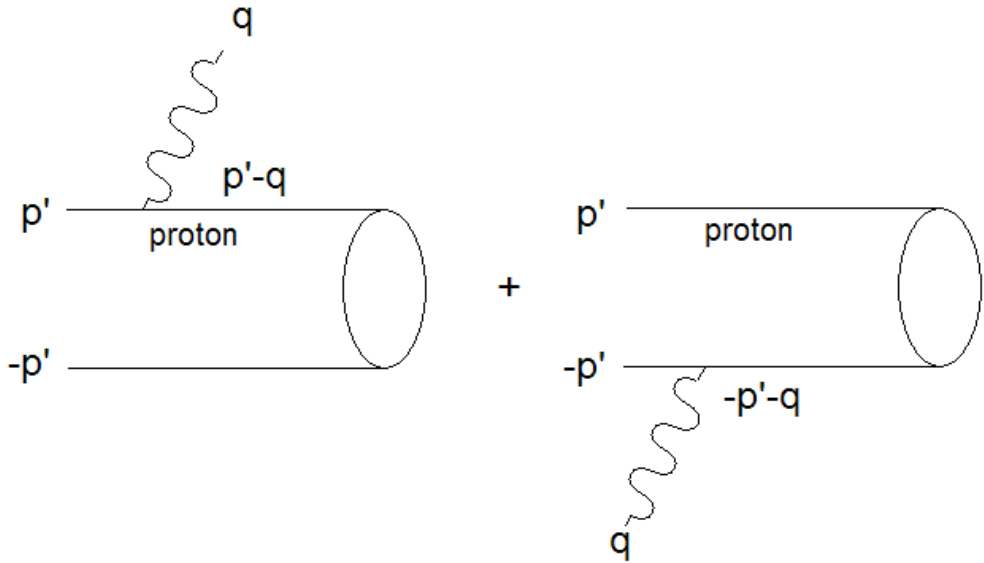}
\end{center}
\par
\vspace{3mm}
\caption{The plane-wave impulse approximation diagrams for deuteron
electro-disintegration. The blob on the right of each diagram represents the
deuteron state. All momenta in the figure are measured in the final
proton-neutron c.m. frame.}
\label{dd}
\end{figure}

The first term on the right-hand side of Eq.~(\ref{eq:6.6a}) is the
plane-wave impulse approximation (PWIA), and the second term is the
final-state interaction (FSI). The dynamics of the PWIA part can be
described by Fig. \ref{dd}: there the final-state proton (neutron) has
3-momentum $\mathbf{p}^{\prime }(-\mathbf{p}^{\prime })$ in the final c.m.
frame, while before the proton (neutron) is struck, it has 3-momentum $%
\mathbf{p}^{\prime }-\mathbf{q}/2$ $(-\mathbf{p}^{\prime }-\mathbf{q}/2)$ in
the deuteron's c.m. frame.

\subsection{Isospin decomposition}

By inserting a complete set of isospin states and using the identity: 
\begin{eqnarray}
\langle \mathbf{p}SM_{s}TM_{T}|G_{0}(E^{\prime })|\mathbf{p}^{\prime \prime
}S^{\prime \prime }M_{s}^{\prime \prime }T^{\prime \prime }M_{T}^{\prime
\prime }\rangle  &=&\delta (\mathbf{p}-\mathbf{p}^{\prime \prime })\delta
_{SS^{\prime \prime }}\delta _{M_{s}M_{s}^{\prime \prime }}\delta
_{TT^{\prime \prime }}\delta _{M_{T}M_{T}^{\prime \prime }}  \notag \\
&&\times \left[ \frac{{\mathcal{P}}}{E^{\prime }-\mathbf{p}^{2}/M}-i\pi
\delta \left( E^{\prime }-\mathbf{p}^{2}/M\right) \right] .
\end{eqnarray}%
we obtain 
\begin{eqnarray}
{\langle \mathbf{p}^{\prime }SM_{s}\Psi _{T}|t(E^{\prime })G_{0}(E^{\prime
})J_{0}(\mathbf{q})|m_{J}0\rangle } =\sum_{T=0,1}\int d^{3}p\,\langle 
\mathbf{p}^{\prime }SM_{s}T|t(E^{\prime })|\mathbf{p}SM_{s}T\rangle   \notag
\\
\times \left[ \frac{{\mathcal{P}}}{E^{\prime }-\mathbf{p}^{2}/M}-i\pi
\delta \left( E^{\prime }-\frac{\mathbf{p}^{2}}{M}\right) \right] \langle 
\mathbf{p}SM_{s}T|J_{0}(\mathbf{q})|m_{J}0\rangle ,  \label{eq:6.10aa}
\end{eqnarray}%
with $E^{\prime }=\frac{\mathbf{p^{\prime }}^{2}}{M},$ and ${\mathcal{P}}$
denotes the principal value. Note that in Eq.~(\ref{eq:6.6a}) the magnitude
of $\mathbf{p}^{\prime }$ is restricted to be that of the proton in the
final-state c.m. frame, while in Eq.~(\ref{eq:6.10aa}) $\mathbf{p}$ is the
integration variable. Note also that in Eq.~(\ref{eq:6.10aa}), and in what
follows, we have dropped the $M_{T}$ label, since all states have $M_{T}=0$:
there are no interactions present that change this quantum number. 

The matrix element $\langle \mathbf{p^{(\prime )}}SM_{s}T|J_{0}(\mathbf{q}%
)|m_{J}0\rangle $ needs to be evaluated so that we can calculate Eq.~(\ref%
{eq:6.6a}). The matrix element can be evaluated as 
\begin{equation}
\langle \mathbf{p}SM_{s}T|J_{0}(\mathbf{q})|m_{J}0\rangle
=\sum_{i=1,2}\langle \mathbf{p}_{i}-\frac{\mathbf{q}}{2}SM_{s}|m_{J}\rangle
(\delta _{T0}G_{E}^{s}(\mathbf{q}^{2})\pm \delta _{T1}G_{E}^{v}(\mathbf{q}%
^{2})),  \label{eq:6.13}
\end{equation}%
where the $+(-)$ sign applies to the $i=1(2)$ case, and the second factor on
each line is now purely a matrix element in isospin space. Here $\mathbf{p}%
_{1(2)}$ is the momentum of particle 1(2) in the final c.m. frame. Note that
particle 1(2) can be either a proton or neutron.

Substituting\bigskip\ Eqs.~(\ref{form}) and (\ref{eq:6.13}) into Eq.~(\ref%
{eq:6.6a}) and Eq.~(\ref{eq:6.10aa}), we get the final expression 
\begin{eqnarray}
\lefteqn{\langle \mathbf{p}^{\prime }SM_{s}\Psi _{T}|(1+tG_{0})J_{0}(\mathbf{%
q})|m_{J}0\rangle }  \notag \\
&=&\langle \mathbf{p}^{\prime }-\frac{\mathbf{q}}{2}\;SM_{s}|m_{J}\rangle
G_{E}^{(p)}(\mathbf{q}^{2})+\langle -\mathbf{p}^{\prime }-\frac{\mathbf{q}}{2%
}\;SM_{s}|m_{J}\rangle G_{E}^{(n)}(\mathbf{q}^{2})  \notag \\
&\quad&+ \frac{1}{2}\sum_{T=0,1} \int d^{3}p \, \langle \mathbf{p}^{\prime
}SM_{s}T|t|\mathbf{p}SM_{s}T\rangle\left[ \frac{{\mathcal{P}}}{E^{\prime }-%
\mathbf{p}^{2}/M}-i\pi \delta (E^{\prime }-\mathbf{p}^{2}/M)\right]  \notag
\\
&&\quad \times \left[ \langle \mathbf{p}-\frac{\mathbf{q}}{2}%
SM_{s}|m_{J}\rangle + (-1)^T \langle -\mathbf{p}-\frac{\mathbf{q}}{2}%
SM_{s}|m_{J}\rangle\right](G_{E}^{(p)}(\mathbf{q}^{2}) + (-1)^T G_{E}^{(n)}(%
\mathbf{q}^{2})).  \label{eq:6.14}
\end{eqnarray}

Note that the final-state interaction piece in Eq.~(\ref{eq:6.14}) itself
has two parts: one for $T=0$ and one for $T=1$. Each part consists of a
t-matrix, the free Green's function and the deuteron wave function with the
nucleon form factors. Here we need to integrate over $\mathbf{p}$, thus the
deuteron wave function is multiplied by the half-shell t-matrix $t$($\mathbf{%
p}^{\prime }\mathbf{,p};E_{np}$). In principle, arbitrarily high values of $|%
\mathbf{p}|$ contribute to the integration which yields the
electrodisintegration amplitude. Physically, this means that the virtual
photon can strike one of the nucleons in a state with arbitrarily large
3-momentum $-\mathbf{q}+\mathbf{p}$ (the other nucleon will have momentum $-%
\mathbf{p}$ in the final proton-neutron c.m. frame). The two nucleons then exchange
momentum to reach their final state through the FSI. This is in contrast
with the PWIA, where, to reach a given final state, the virtual photon must
strike the nucleon at a specific momentum. However, in practice, the high-$|%
\mathbf{p}|$ component of the deuteron wave function is small, so the
high-momentum part of the FSI integral will be suppressed.

At this point, we have an expression for the sum of a
plane-wave-impulse-approximation piece and the final-state interaction in
terms of the 3-momentum of the measured proton $\mathbf{p}^{\prime }$. The
next step is to express Eq.~(\ref{eq:6.14}) in terms of partial waves.

\subsection{Partial-wave decomposition}

\label{sec-d3}

\subsubsection{Plane-wave impulse approximation}

\begin{figure}[tbp]
\begin{center}
\includegraphics[width=14cm]{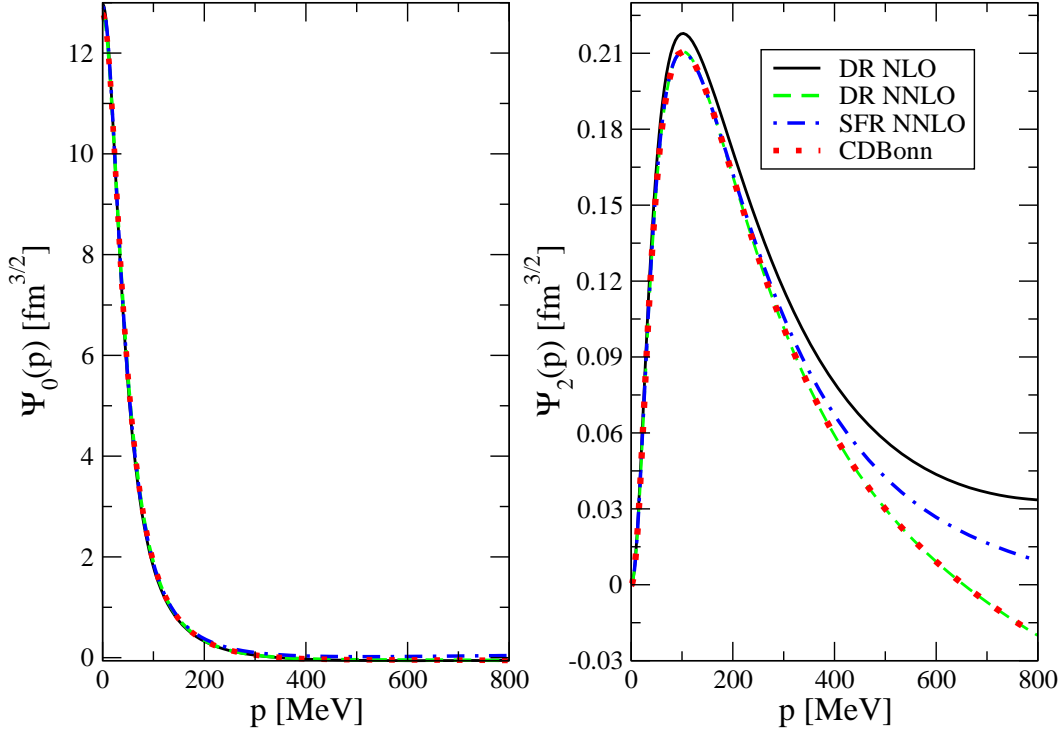}
\end{center}
\par
\vspace{3mm}
\caption{ The momentum-space wave functions of the deuteron as a function of
p, where $\protect\psi _{0}$(p) is the $^{3}$S$_{1}$ wave function and $%
\protect\psi _{2}$(p) denotes the $^{3}$D$_{1}$ wave function. These wave
functions are obtained from the dimensional-regularization (DR)
two-pion-exchange potential (TPE) up to next-to leading order (NLO), i.e., $%
\mathcal{O}(Q^2)$ (black solid line), DR TPE up to next-to-next-to leading
order (NNLO), i.e., $\mathcal{O}(Q^3)$ (green dashed line), and the
spectral-function-regularization (SFR) TPE up to NNLO (blue dash-dotted
line), where the cutoff in the Lippmann-Schwinger equation (LSE) is $\Lambda
=800$ MeV, and the SFR intrinsic cutoff is $\widetilde{\Lambda }=800$ MeV.
The red dots indicate the corresponding wave functions obtained from the
CD-Bonn potential. }
\label{drnlow}
\end{figure}

First, we perform the partial-wave decomposition of the PWIA part of Eq.~(%
\ref{eq:6.14}). To do this we insert 
\begin{equation}
1=\frac{2}{\pi }\sum_{\substack{ J,m_{J},LS  \\ with|L-S|\leq J\leq |L+S|}}%
\int\limits_{0}^{\infty }dp \, p^{2}|pJm_{J}LS\rangle \langle pJm_{J}LS|
\label{eq:6one}
\end{equation}%
into $\langle \mathbf{p}^{\prime} -\frac{\mathbf{q}}{2} SM_{s}|m_{J}\rangle $%
, where $p=|\mathbf{p}|$. Note that the normalization adopted here is%
\begin{equation}
\langle \mathbf{k}^{\prime }|\mathbf{k}\rangle =\delta(\mathbf{k}-\mathbf{k}%
^{\prime })
\end{equation}%
and hence 
\begin{equation}
\langle p^{\prime }J^{\prime }m_{J^{\prime }}L^{\prime }S^{\prime
}|pJm_{J}LS\rangle =\frac{\pi }{2}\frac{\delta (p-p^{\prime})}{p^{2}}\delta
_{JJ^{\prime }}\delta_{m_{J}m_{J^{\prime }}} \delta _{L^{\prime }L}\delta
_{S^{\prime }S}.  \label{eq:6.19}
\end{equation}%
Meanwhile, the deuteron wave function appears as matrix elements 
\begin{equation}
\langle |\mathbf{p}^{\prime }-\frac{\mathbf{q}}{2} \mathbf{|} \,
Jm_{J}LS|m_{J}\rangle =i^{L}\Psi _L\left(|\mathbf{p}^{\prime }- \frac{%
\mathbf{q}}{2}\mathbf{|}\right),  \label{eq:6.16}
\end{equation}
where the S- ($L=0$) and D- ($L=2$) wave components of the deuteron wave
function satisfy%
\begin{equation}
\frac{2}{\pi }\int dpp^{2}[\Psi _{0}^{2}(p)+\Psi _{2}^{2}(p)]=1.
\end{equation}
The momentum-space wave functions we employ are obtained using the methods
discussed in ref.\cite{Ya08}, and are shown, together with those of the
CD-Bonn potential, in Fig.~\ref{drnlow}.

\bigskip 

Putting this all together and using the plane-wave expansion formula (with
S=1 for deuteron) 
\begin{equation}
|\mathbf{p}^{\prime } 1 M_{s}\rangle =\sqrt{\frac{2}{\pi }}\sum_{L^{\prime
},m_{l}^{\prime }}Y_{L^{\prime }m_{l}^{\prime }}^{\ast }(\Omega _{p^{\prime
}})|p^{\prime}L^{\prime }m_{l}^{\prime }\rangle \otimes |1 M_{s}\rangle ,
\label{eq:6.18}
\end{equation}
the two contributions to the PWIA matrix element become: 
\begin{eqnarray}  \label{eq:6.20}
\langle \pm \mathbf{p}^{\prime }-\frac{\mathbf{q}}{2} \; 1
M_{s}|m_{J}\rangle &=&\sqrt{\frac{2}{\pi }}\sum_{L=0,2}\sum_{m_l}(Lm_{l}
1M_{s}|1m_{J}) Y_{Lm_{l}}(\Omega _{\pm p^{\prime }-q/2}) i^L \Psi _{L}\left(%
\mathbf{|} \mathbf{p}^{\prime }\mp\frac{\mathbf{q}}{2} \mathbf{|}\right). 
\notag \\
\end{eqnarray}
Here $(Lm_{l} 1M_{s}|1m_{J})$ is the Clebsch-Gordan coefficient and $\Omega
_{\pm p^{\prime }-q/2}$ is the angle between $\widehat{z}$ and $\pm \mathbf{p%
}^{\prime }-\frac{\mathbf{q}}{2}$.

\subsubsection{Final-state interaction}

One can follow the same logic as for the PWIA piece to perform the
partial-wave decomposition of the FSI term. Note that the total spin ($S$)
and isospin ($T$) are conserved in the NN interaction. Thus 
\begin{eqnarray}  \label{eq:FSIme}
&&\langle \mathbf{p}^{\prime }SM_{s} T|t(E^{\prime })G_{0}(E^{\prime })J_0(%
\mathbf{q})|m_{J} 0\rangle  \notag \\
&&=\sum_{M_{s}^{\prime }}\int d^{3}p \langle \mathbf{p}^{\prime }SM_{s}
T|t(E^{\prime })|\mathbf{p}SM_{s}^{\prime } T\rangle \left[ \frac{{\mathcal{P%
}}}{E^{\prime }-\mathbf{p}^{2}/M}-i\pi \delta \left(E^{\prime }-\frac{%
\mathbf{p}^{2}}{M}\right)\right]\langle \mathbf{p}SM_{s}^{\prime } T|J_0(%
\mathbf{q})|m_{J} 0\rangle .  \notag \\
\end{eqnarray}
The matrix element $\langle \mathbf{p}S^{\prime }M_{s}^{\prime }T|J_0(%
\mathbf{q})|m_{J}0\rangle $ is evaluated via the first line of Eq.~(\ref%
{eq:6.14}) and Eq.~(\ref{eq:6.20}). The 3-dimensional t-matrix $t(\mathbf{p}%
^{\prime },\mathbf{p};E^{\prime })$ can be constructed from $%
t_{LSJ}(p^{\prime },p;E)$. Since the deuteron has $S=1$, we need only $%
t_{L1J}(p^{\prime },p;E)$, for which: 
\begin{eqnarray}
&&\langle \mathbf{p}^{\prime } 1 M_{s}T_{f}|t(E^{\prime })|\mathbf{p}1
M_{s}^{\prime }T_{f}\rangle  \notag \\
&&\qquad=\frac{2}{\pi } \sum_{_{\substack{ J,m_{j},L,L^{\prime \prime }  \\ %
m_{l}m_{l}^{\prime \prime }}}}(Lm_{l}1M_{s}|Jm_{j})Y_{Lm_{l}}(\Omega
_{p^{\prime }})t_{L1J,T_{f}}(E^{\prime })(L^{\prime \prime }m_{l}^{\prime
\prime }1M_{s}^{\prime }|Jm_{j})Y_{L^{\prime \prime }m_{l}^{\prime \prime
}}^{\ast }(\Omega _{p}).  \label{eq:6.26}
\end{eqnarray}

Upon insertion of these results into Eq.~(\ref{eq:FSIme}) we find that we
need to perform the angular part of the integral in the following form%
\begin{equation}
\int Y_{L^{\prime \prime }m_{l}^{\prime \prime }}^{\ast }(\Omega
_{p})Y_{Lm_{l}}(\Omega _{\pm p-q/2})d\Omega _{p}.
\end{equation}%
Here $\Omega _{\pm p-q/2}$ is the solid angle between $\widehat{z}$ and $\pm 
\mathbf{p-q/}2$. We now denote the angle between $\widehat{z}$ and $\mathbf{%
p-q/}2$ as ($\theta ^{\prime },\phi $), and%
\begin{equation}
\theta ^{\prime }=\sin ^{-1}\left[\frac{|\mathbf{p}|\sin \theta }{\sqrt{|%
\mathbf{p}|^{2}-|\mathbf{p}||\mathbf{q}|\cos \theta +|\mathbf{q}|^2/4}}%
\right],  \label{eq:4}
\end{equation}%
with $(\theta,\phi)$ the solid angle between $\mathbf{p}^{\prime}$ and $\hat{z}$.
Similarly, the angle between $\widehat{z}$ and $-\mathbf{p-q/}2$, which we
denote as ($\theta ^{\prime \prime },\phi +\pi $), is%
\begin{equation}
\theta ^{\prime \prime }=\sin ^{-1}\left[\frac{|\mathbf{p}|\sin \theta }{%
\sqrt{|\mathbf{p}|^{2}+|\mathbf{p}||\mathbf{q}|\cos \theta +|\mathbf{q}|^2/4}%
}\right].  \label{eq:5}
\end{equation}

Thus, to evaluate $\langle \mathbf{p}^{\prime }S=1M_{s}T_{f}|t(E^{\prime
})G_{0}(E^{\prime })J_0(\mathbf{q})|m_{J}0\rangle ,$ we need to do the
integral over the angles $\theta $ and $\phi $. The integral over $\phi $
can be reduced by taking advantage of the property of spherical harmonics:%
\begin{eqnarray}
&&\int_{0}^{2\pi }\int_{0}^{\pi }Y_{lm}^{\ast }(\theta ,\phi )Y_{l^{\prime
}m^{\prime }}(\theta ^{\prime },\phi )d(\cos \theta )d\phi  \notag \\
&=&2\pi \delta _{mm^{\prime }}\sqrt{\frac{2l+1}{4\pi }\frac{(l-m)!}{(l+m)!}%
\frac{2l^{\prime }+1}{4\pi }\frac{(l^{\prime }-m)!}{(l^{\prime }+m)!}}%
\int_{0}^{\pi }P_{l}^{m}(\cos \theta )P_{l^{\prime }}^{m}(\cos \theta
^{\prime })d(\cos \theta ).  \label{eq:6}
\end{eqnarray}
With the aid of Eq.~(\ref{eq:6}), we can then perform the numerical
integration over $\theta $ and $p$ to obtain the final-state interaction
contribution to the longitudinal response function.

\begin{figure}[tbp]
\begin{center}
\includegraphics[width=14cm]{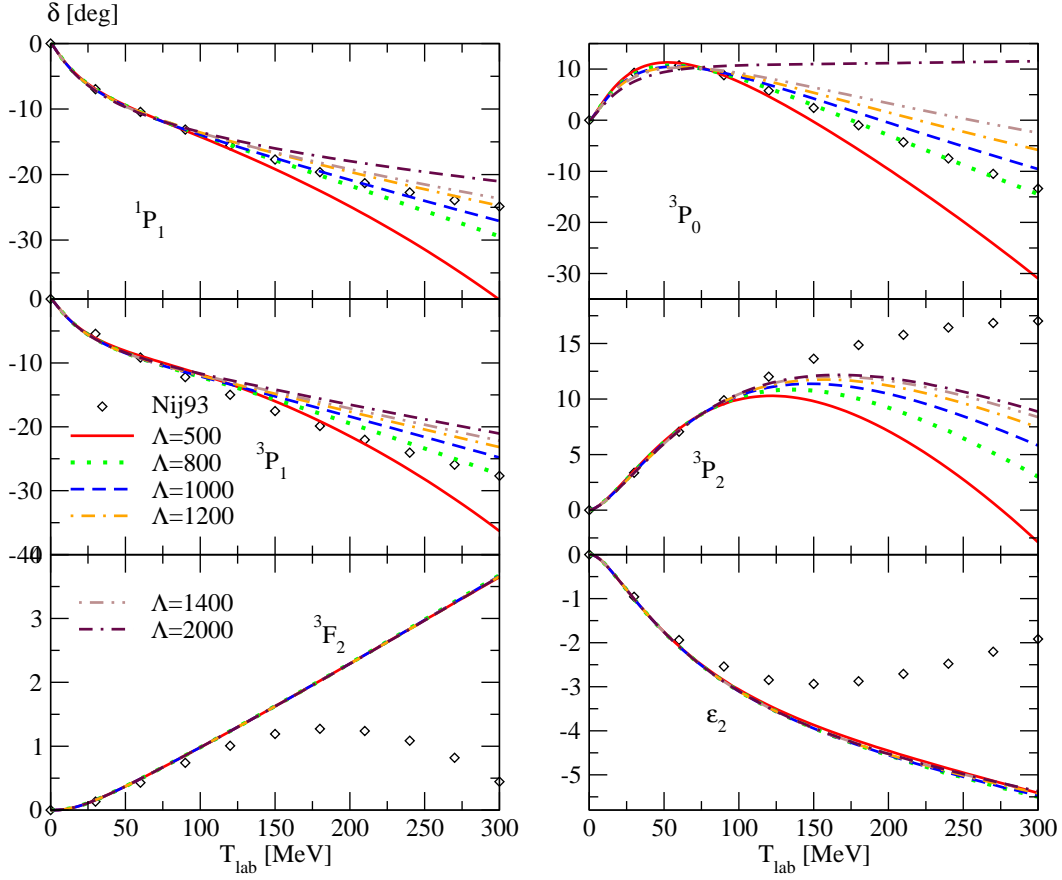}
\end{center}
\par
\vspace{3mm}
\caption{ The NN P-wave phase shifts as a function of the laboratory kinetic
energy that result from choosing $v^{LR}=v_{1\protect\pi }+v_{2\protect\pi }$%
, with the latter chosen to be the SFR TPE up to NNLO (with intrinsic cutoff 
$\widetilde{\Lambda }=800$ MeV). Here the generalized scattering lengths $%
\protect\alpha _{11}^{SJ}$ are adjusted to give the best fit in the region $%
T_{lab}<100$~MeV. The Nijmegen phase-shift analysis~ \protect\cite%
{nnonline,St93} is indicated by the open diamonds. This graph is adapted
from our previous publication \protect\cite{Ya09A}.}
\label{figfullbest}
\end{figure}

\begin{figure}[tbp]
\begin{center}
\includegraphics[width=10cm]{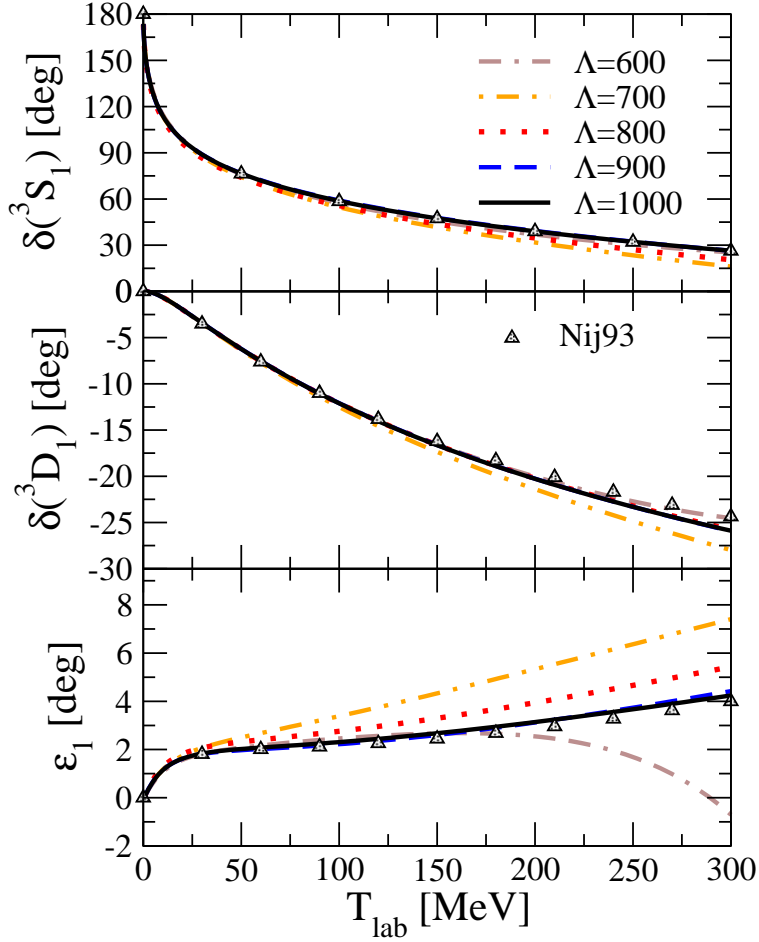}
\end{center}
\par
\vspace{3mm}
\caption{ The best fit for the NN $^{3}$S$_{1}-^{3}$D$_{1}$ phase shifts as
a function of the laboratory kinetic energy for different cutoffs $\Lambda $
ranging from 0.6 to 1~GeV. The potentials employed are the SFR NNLO (with
intrinsic cutoff $\widetilde{\Lambda }=800$ MeV). The values of the Nijmegen
phase-shifts~ \protect\cite{nnonline,St93} are indicated by the open
triangles. }
\label{fig-fig200}
\end{figure}

In doing this it is clearly important to have a description of the NN
interaction that agrees with data for NN final-state energies of interest.
In fact, the LECs of the NN t-matrix we adopted in our calculation of $f_L$
are those which generate phase shifts that agree with the Nijmegen
phase-shift analysis~\cite{nnonline,St93} for $T_{lab}\leq 100$ MeV (as can
be seen in Figs.~\ref{figfullbest} and \ref{fig-fig200}). This kinematics
corresponds to $E_{np}\leq 50$ MeV.


\section{Results for the longitudinal response function}

\label{sec-d4}

In this section we present our results for the longitudinal response
function of deuteron electro-disintegration. For the nucleon form factors we
adopt the results listed in Ref. \cite{Bel}. For electro-disintegration, one
needs to specify two kinematic variables, e.g., ($\omega ,\mathbf{q}$) to
describe the whole process. We adopted the following kinematic variables in
order to compare our results with those obtained with the Bonn potential in
Ref. \cite{aren}: first, the final energy of the proton-neutron system,
which hereafter is labeled $E_{np},$ (previously it was denoted $E^{\prime}$%
) i.e.,%
\begin{equation}
E_{np}=\frac{\mathbf{p}^{\prime 2}}{M};  \label{eq:6.00}
\end{equation}%
second, the 3-momentum of the virtual photon (also in the system's final
c.m. frame) $\mathbf{q}_{cm}^{2}$. With $E_{np}$ and $\mathbf{q}_{cm}^{2}$
specified, the energy of the virtual photon can be calculated to be%
\begin{equation}
\omega_{cm} =E_{np}-\sqrt{M_{d}^{2}+\mathbf{q}_{cm}^{2}}+2M.  \label{eq:6.28}
\end{equation}

The experimental data of Refs.~\cite{vdS91,Du94,Jo96} are presented in terms
of the lab. frame value of $|\mathbf{q}|$ and the value of the ``missing
momentum", 
\begin{equation}
\mathbf{p}_m=\mathbf{p}^{\prime}_{lab}-\mathbf{q}_{lab},
\end{equation}
with $\mathbf{p}_{lab}^{\prime}$ the momentum of the detected proton. All
these data were taken in kinematics such that $\mathbf{p}_{lab}^{\prime}$
and $\mathbf{q}_{lab}$ are aligned, and so $\theta=\phi=0$. From this
information, and knowledge of the virtual-photon energy, $\omega_{lab}$, we
can compute the kinetic energy of the np pair in the lab frame in two
different ways: 
\begin{equation}
E_{np,lab}=\omega_{lab} + M_d-2M=\sqrt{M^2 + \mathbf{p}_m^2} + \sqrt{M^2 + 
\mathbf{p}^{\prime 2}} - 2 M.
\end{equation}
Lorentz transformation of this quantity to the cm frame according to \ 
\begin{equation}
E_{np,cm}=\gamma(E_{np,lab} - \beta |\mathbf{{q}|),}
\end{equation}
with $\beta$ given by Eq.~(\ref{eq:beta}), yields the $E_{np}$ which we
quote in our results. Alternatively, $E_{np,cm}$ can be obtained by energy
conservation, applied in the cm frame: 
\begin{equation}
E_{np,cm}=\sqrt{M_d^2 + \mathbf{q}_{cm}^2} + \omega_{cm} - 2 M
\end{equation}
where $\omega_{cm}$ and $\mathbf{q}_{cm}$ are obtained from $\omega_{lab}$
and $\mathbf{q}_{lab}$ using Eq.~(\ref{eq:LT}).

Before presenting the results, we introduce one kinematics which is of
particular interest: the so-called quasi-free ridge. The quasi-free ridge
occurs when $\omega _{cm}=0$. Physically, this means that the virtual photon
hits one of the nucleons and gives just enough 3-momentum to put it
on-mass-shell. The other nucleon remains at rest in the laboratory frame.
This occurs when: 
\begin{equation}
E_{np,cm}\approx M_{d}(1+\frac{\mathbf{q}_{cm}^{2}}{2M_{d}^{2}})-2M\approx 
\frac{\mathbf{q}_{cm}^{2}}{2M_{d}}.  \label{eq:6.29}
\end{equation}%
Consequently, on the quasi-free ridge $E_{np}$ (in MeV)$\approx 10\,\mathbf{q%
}_{cm}^{2}$ (in fm$^{-2}$).

\begin{figure}[tbp]
\begin{center}
\includegraphics[width=14cm]{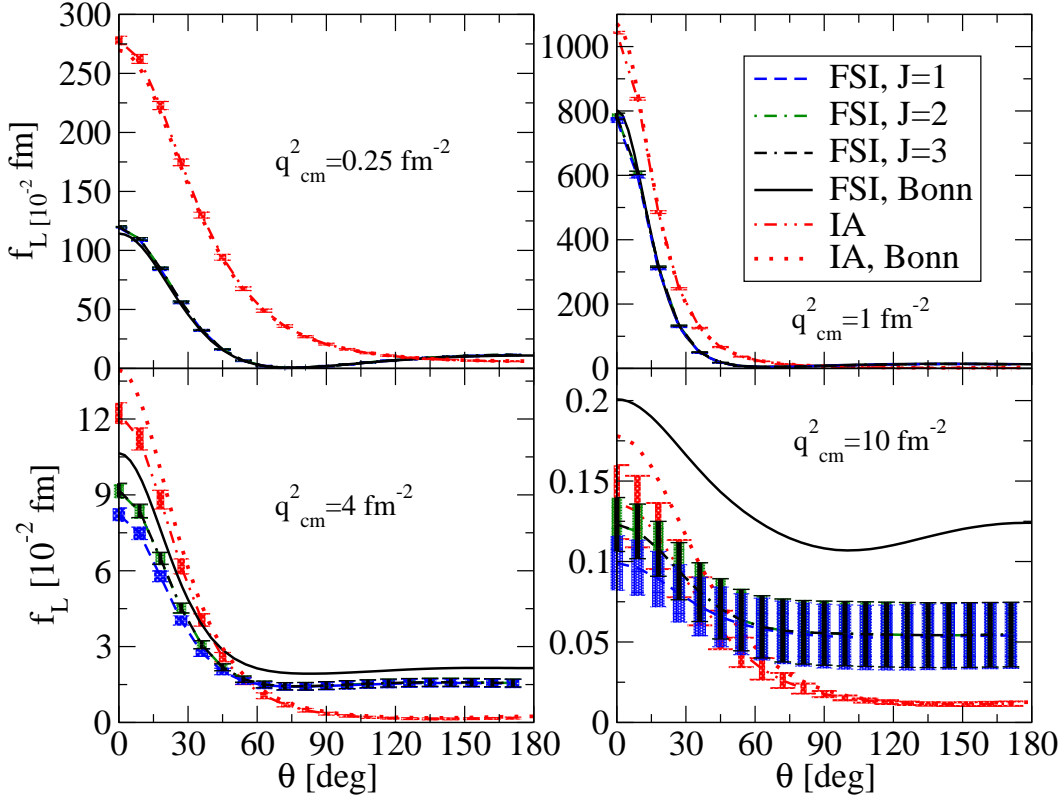}
\end{center}
\par
\vspace{3mm}
\caption{ The deuteron electro-disintegration longitudinal response function 
$f_{L}$ as a function of $\protect\theta $ for $E_{np}=10$ MeV and four
different values of $\mathbf{q}_{cm}^{2}$. The error bars are obtained from
varying the cutoff $\Lambda =600-1000$ MeV in the LSE. The convergence of
the calculation as additional partial waves are included in the computation
of the FSI is also shown. Legend as indicated. Here the intrinsic cutoff of
SFR TPE is $\widetilde{\Lambda }=800$ MeV.}
\label{figex}
\end{figure}

We now present our results. First, we adopt the NN t-matrix and deuteron
wave function generated with spectral-function regularization (SFR) applied
to the two-pion-exchange potential up to NNLO, with the SFR cutoff $%
\widetilde{\Lambda }$ set to $800$ MeV. Fig.~\ref{figex} shows the
longitudinal structure function $f_{L}$ versus angle $\theta $, i.e., the
angle between $\mathbf{p}^{\prime }$ and $\mathbf{q,}$ for $E_{np}=10$ MeV
and $\mathbf{q}_{cm}^{2}=0.25-25$ fm$^{-2}$. The $\chi$EFT PWIA result is
denoted by the red dash-double-dotted line, with error bars indicating the
effect of varying the cutoff in the Lippmann-Schwinger equation (LSE) from $%
\Lambda =600-1000$ MeV. It is obtained using the full deuteron wave function
obtained from the SFR TPE potential, evaluated at the pertinent
three-momentum, see Eq.~(\ref{eq:6.20}).

For the final-state interaction (FSI), one needs to sum over partial-waves
in order to obtain the 3-dimensional t-matrix. We have summed over
partial-waves up to $J=3,$ and the results are denoted as blue dashed ($J=1$%
), green dash-dotted ($J$ up to $2$) and black double-dash-dotted ($J$ up to 
$3$) line in Fig.~\ref{figex}\footnote{%
Note that for $J\geq 2$, there is no contact term associated with the long
range part of the TPE in the standard Weinberg's power counting. Here we
adopt the Born approximation instead of iterating the SFR\ TPE in the LSE.}.
In general, our results converge once we include partial waves with $J=2$ in
our calculation of the FSI.

Here and below we compare our calculations to the calculations of
Arenh\"ovel and collaborators~\cite{aren,Aren10}. These calculations are
done using the Bonn-B potential~\cite{MHE87} and include PWIA and FSI
pieces. This allows a direct comparison with the PWIA and FSI-included $\chi$%
EFT calculations that are the subject of this work. The PWIA result obtained
with the Bonn-potential wave function is indicated by the red dotted line.
When both $E_{np}$ and $\mathbf{q}_{cm}^{2}$ are low, i.e., $E_{np}=10$ MeV
and $\mathbf{q}_{cm}^{2}\leq 1$ fm$^{-2}$, the results agree very well. As $%
\mathbf{q}_{cm}^{2}$ becomes larger, the PWIA\ results obtained from the two
potentials start to deviate from each other.

We first discuss the quasi-free ridge case shown in Fig.~\ref{figex}, i.e., $%
E_{np}=10$ MeV and $\mathbf{q}_{cm}^{2}=1$ fm$^{-2}$. Out of all four panels
in Fig.~\ref{figex}, $f_{L}$ receives the least correction from the FSI
here, and, as shown in the other three panels, the further away we move from
the quasi-free ridge the larger the FSI correction becomes. This can be
explained easily by the fact that, at the quasi-free ridge, both nucleons in
the deuteron are on the mass shell after being struck by the virtual photon,
and no FSI is needed in order to make the final-state particles real. On the
other hand, as we move further kinematically from the quasi-free ridge, the
FSI must provide a larger energy-momentum transfer to make the proton and
neutron become real particles in the final state, and so it becomes more
important.

Moreover, for this particular np final-state energy, the quasi-free ridge is
the last $\mathbf{q}_{cm}^2$ where all the PWIA and FSI results from the two
potentials agree. As we increase $\mathbf{q}_{cm}^{2}$ to 4 fm$^{-2}$ and
above, both our PWIA and FSI results start to diverge away from the
corresponding Bonn potential results. The error bars also grow quite
significantly for $\mathbf{q}_{cm}^{2} > 4$ fm$^{-2}$, i.e. $\mathbf{q}_{cm}
> 400$ MeV, particularly in the FSI. There, where both the deuteron wave
functions and the NN t-matrix enter the calculation, the results become
highly cutoff-dependent.

\begin{figure}[tbp]
\begin{center}
\includegraphics[width=14cm]{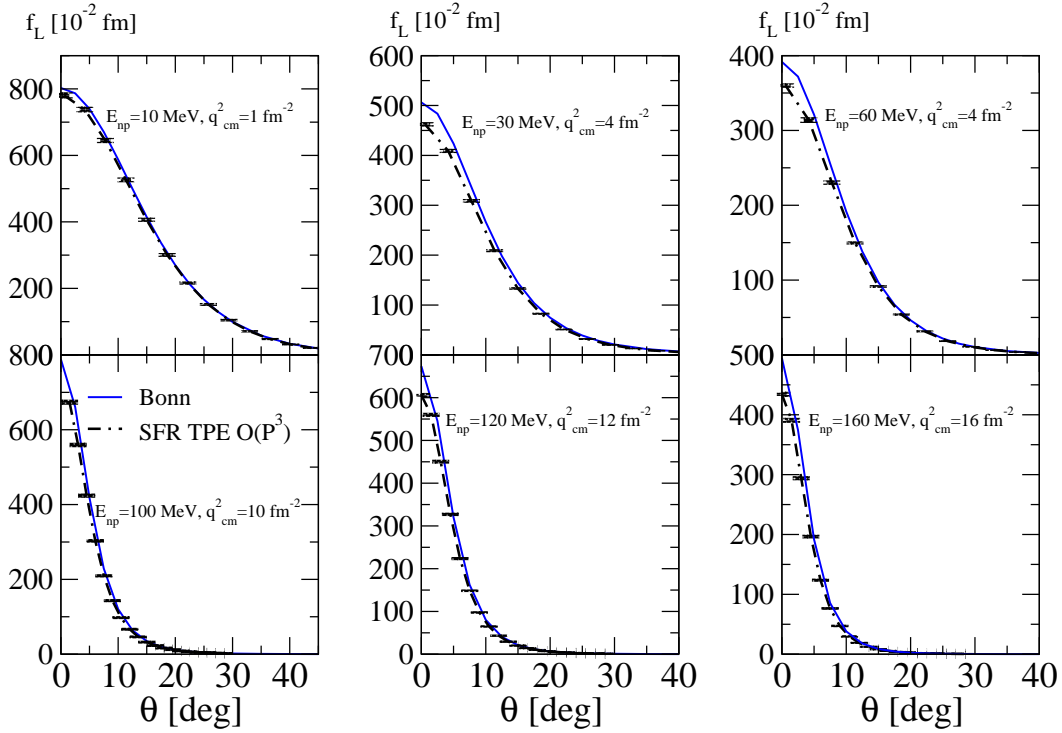}
\end{center}
\par
\vspace{3mm}
\caption{The deuteron electro-disintegration longitudinal response function $%
f_{L}$ as a function of $\protect\theta $ in various kinematics on or around
the ``quasi-free ridge". Here the blue line represents the result given by
the Bonn potential with FSI included, and the black double-dash-dotted line
denotes the same results given by the SFR TPE up to NNLO. The error bars are
obtained from varying the cutoff $\Lambda =600-1000$ MeV in the LSE. The SFR
intrinsic cutoff is $\widetilde{\Lambda }=800$ MeV.}
\label{quasi}
\end{figure}

We now assess how this uncertainty in the $\chi$EFT $f_{L}$ prediction comes
from the uncertainty of the $\chi$EFT deuteron wave function and NN
t-matrix. Let's first look at the quasi-free ridge. From Eq.~(\ref{eq:6.00})
and Eq.~(\ref{eq:6.29}), we infer that 
\begin{equation}
|\mathbf{p}^{\prime }| \approx \frac{|\mathbf{q}_{cm}|}{2}
\end{equation}%
at the quasi-free ridge, where the dominant element in the calculation is
the deuteron wave function $\Psi _{L}(|\mathbf{p}^{\prime }-\frac{\mathbf{q}%
_{cm}}{2}\mathbf{|})$. At the quasi-free ridge, the value of the
wave-function argument achieves its lowest possible value (for a given $%
E_{np}$): it is $0$ (for $\theta =0$) increasing to $2|\mathbf{p}^{\prime }|$
(for $\theta =\pi $). Fig.~\ref{drnlow} shows that the deuteron wave
function $\Psi _{L}(p\mathbf{)}$ given by both the SFR and DR\ TPE\ up to
NNLO agrees with the one given by the Bonn potential at least up to wave
function arguments $\approx 100$ MeV, and dies off quickly at higher
momentum. Since the high-momentum component of the wavefunction is almost
zero\footnote{%
The $^{3}D_{1}$ wavefunction in momentum-space $\Psi _{2}(p\mathbf{)}$ dies
off at a higher momentum, i.e., $p>400$ MeV. However, it is at least 10
times smaller in amplitude than the $^{3}S_{1}$ wavefunction $\Psi _{0}(p%
\mathbf{)}$.}, this suggests that $f_{L}$ calculated from these two
potentials should agree with each other at the quasi-free ridge. In fact, as
shown in Fig.~\ref{quasi}, in quasi-free kinematics the $f_{L}$ given by the
SFR\ TPE up to NNLO does agree with those given by the Bonn potential all
the way up to $E_{np}=160$ MeV.

To see where the two wave functions start to disagree, we use Fig.~\ref%
{plot_en10} as an example ($E_{np}=10$ MeV again, now at more $\mathbf{q}%
_{cm}^2$ values). For $\mathbf{q}_{cm}^2 \le 1$ fm$^{-2}$, there is no
significant difference between results obtained by $\chi$EFT and the Bonn
potential. At $\mathbf{q}_{cm}^2=4$ fm$^{-2}$, the shift due to FSI is
roughly the same ($\approx 3.7\times 10^{-2}$ fm at $\theta =0$) for both
the NNLO SFR TPE and the Bonn potential. In other words, the FSI has almost
the same effect for the two potentials, and the disagreement in the total $%
f_L$ comes (mostly) from the PWIA part. The $\approx 10\% $ difference in
the PWIA amplitude originates from the difference between the deuteron wave
functions generated by NNLO SFR\ TPE and the Bonn potentials at around $%
p=100 $ MeV and is not significantly enlarged by the FSI piece where the
deuteron wave function is integrated against the NN t-matrix.

\begin{figure}[tbp]
\begin{center}
\includegraphics[width=14cm]{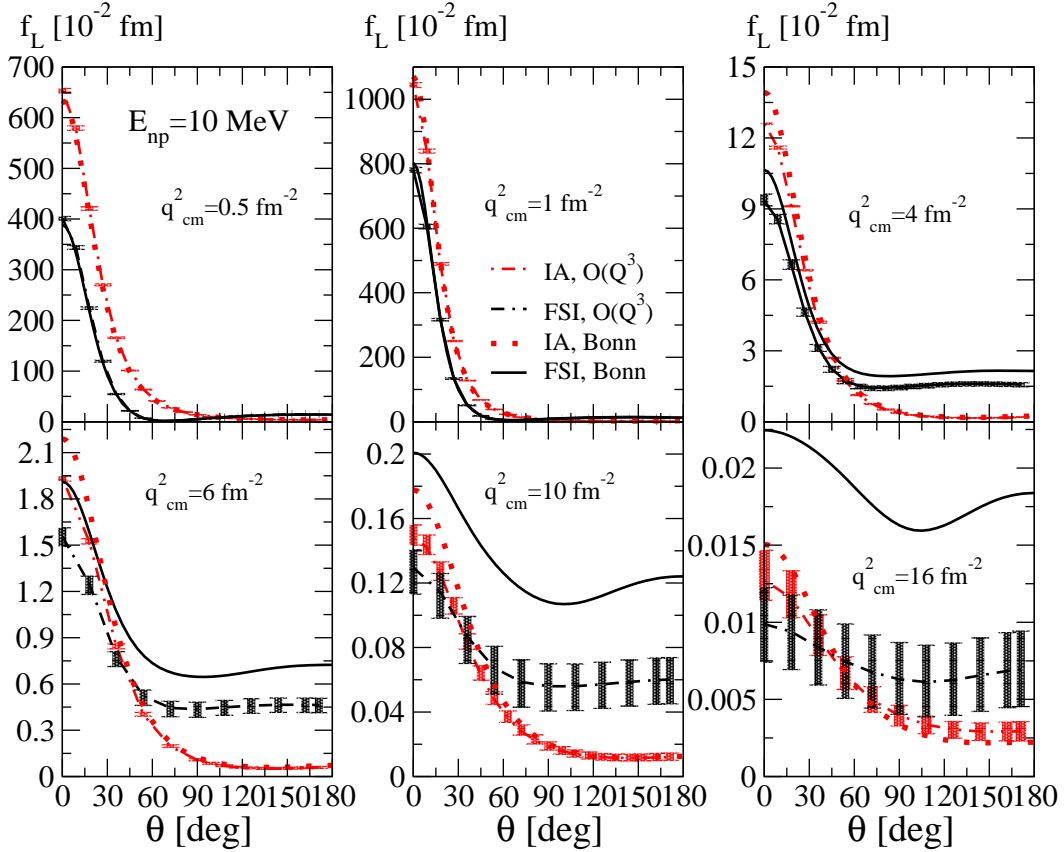}
\end{center}
\par
\vspace{3mm}
\caption{ The deuteron electro-disintegration longitudinal response function 
$f_{L}$ as a function of $\protect\theta $ for $E_{np}=10$ MeV, and $\mathbf{%
q}^2_{cm}$ ranging from 0.5 fm$^{-2}$ to 16 fm$^{-2}$. Here the red
dash-double-dotted (dotted) line represents the PWIA results given by the
SFR TPE up to NNLO (Bonn), and the black double-dash-dotted (solid) line
denotes the FSI results given by the SFR TPE up to NNLO (Bonn). The error
bars are obtained from varying the cutoff $\Lambda =600-1000$ MeV in the
LSE. The intrinsic cutoff is $\widetilde{\Lambda }=800$ MeV for the SFR TPE
up to NNLO.}
\label{plot_en10}
\end{figure}

On the other hand, as we increase $\mathbf{q}_{cm}^{2}$ to $10$ fm$^{-2}$
(with $E_{np}=10$ MeV) $f_{L}$ given by the two different potentials starts
to have a larger difference in the FSI than in the PWIA---see the lower
panels of Fig.~\ref{plot_en10}. Although this final-state energy is well
within the range that is fit by our NN potential, one must remember that the
deuteron wave function that enters the FSI integral is largest when $|%
\mathbf{p}|=|\mathbf{q}_{cm}|/2$, and the phase-shift data---where we
perform best fit up to $T_{lab}=100$ MeV---only validates our computation of 
$t(p^{\prime },p;E_{np})$ for $E_{np} \le 50$ MeV and $p$ up to about $225$
MeV. We infer that it is important for $t(p^{\prime },p;E_{np})$ to at least
accurately describe data for the on-shell kinematics corresponding to both $%
p^{\prime }(\equiv \sqrt{ME_{np}})$ and $p(= |\mathbf{q}_{cm}|\mathbf{/}2)$.
If either of these is greater than $225$ MeV, then the difference in the NN
t-matrix generated by the SFR\ TPE\ up to NNLO and the Bonn potential enters
the FSI calculation in addition to any differences in $\Psi _{L}$.

\begin{figure}[tbp]
\begin{center}
\includegraphics[width=14cm]{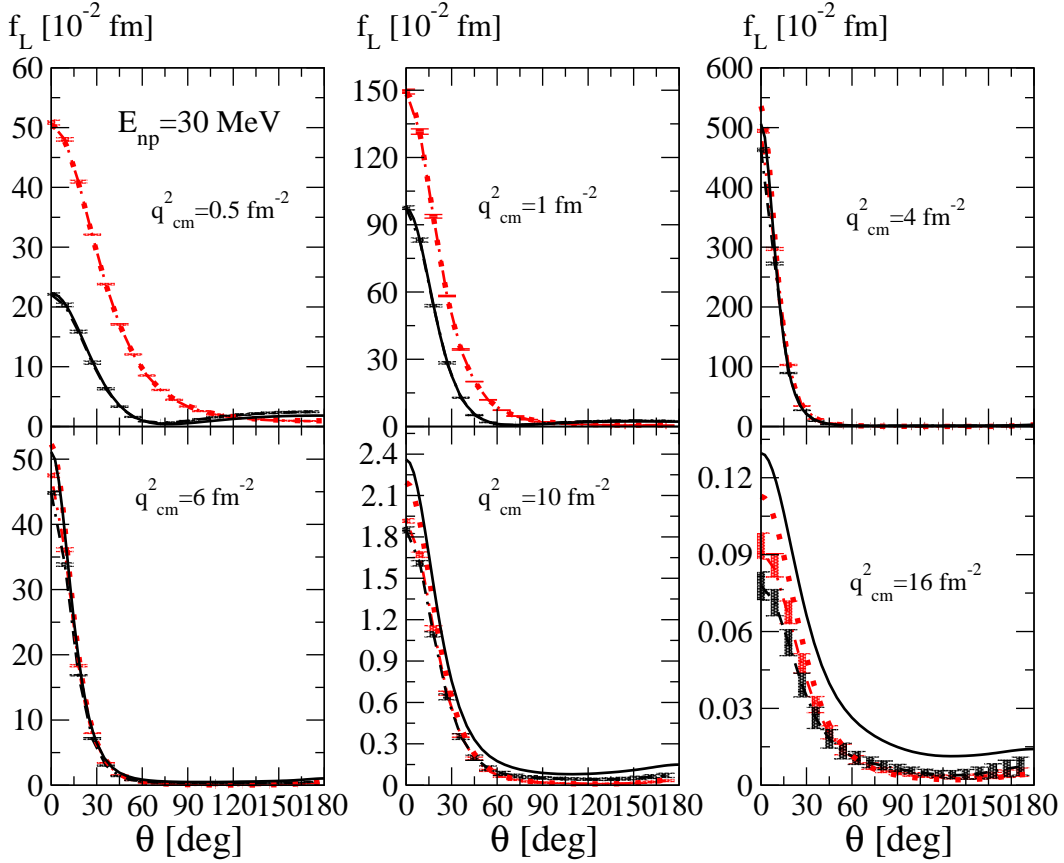}
\end{center}
\par
\vspace{3mm}
\caption{ The deuteron electro-disintegration longitudinal response function 
$f_{L}$ as a function of $\protect\theta $ for $E_{np}=30$ MeV, and $\mathbf{%
q}^2_{cm}$ ranging from 0.5 fm$^{-2}$ to 16 fm$^{-2}$. Legend as in Fig.~%
\protect\ref{plot_en10}.}
\label{plot_en30}
\end{figure}

In Fig.~\ref{plot_en30} we show a similar set of panels to those in Fig.~\ref%
{plot_en10}, but at $E_{np}=30$ MeV. The agreement between $\chi$EFT and
Bonn results is again quite good at low $\mathbf{q}^2$, although there is
some disagreement at backward angles once FSI is included. This trend in the
final result for $f_L$ diminishes as we move towards the quasi-free ridge.
At the quasi-free ridge the cutoff variation of the $\chi$EFT calculation is
small---smaller than for $E_{np}=10$ MeV, because the FSI plays less of a
role at this higher energy. The agreement with the Bonn potential is also
quite good there. Immediately above the quasi-free ridge these features
persist, until $\mathbf{q}^2_{cm} \approx 6$ fm$^{-2}$, at which point $|%
\mathbf{q}_{cm}|$ becomes large enough, and the FSI important enough, that
agreement cannot be maintained.

\begin{figure}[tbp]
\begin{center}
\includegraphics[width=14cm]{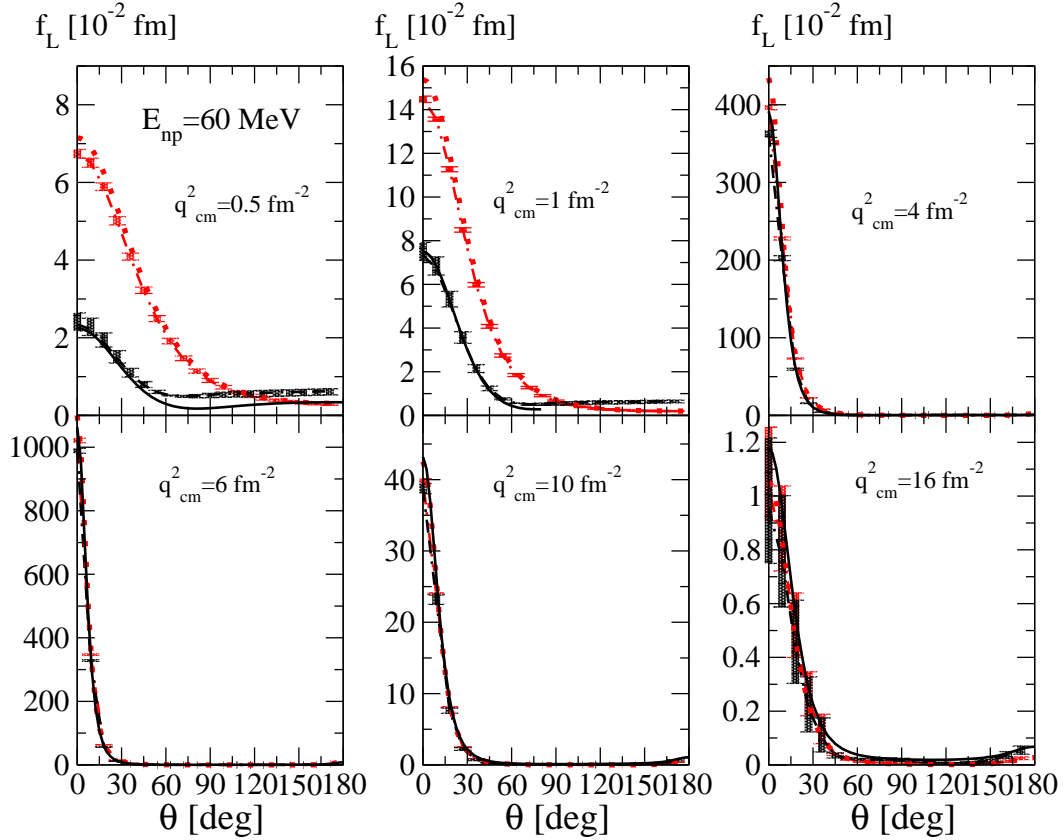}
\end{center}
\par
\vspace{3mm}
\caption{ The deuteron electro-disintegration longitudinal response function 
$f_{L}$ as a function of $\protect\theta $ for $E_{np}=60$ MeV, and $\mathbf{%
q}^2_{cm}$ ranging from 0.5 fm$^{-2}$ to 16 fm$^{-2}$. Legend as in Fig.~%
\protect\ref{plot_en10}.}
\label{plot_en60}
\end{figure}

If we now examine $E_{np}=60$ MeV we are already in a regime where the FSI
is not trustworthy even if $\mathbf{q}^2_{cm}$ is low. This is reflected in
the failure of the FSI-included result to encompass the Bonn-potential
answer, even within error bars, at $\mathbf{q}^2_{cm}=0.5$ fm$^{-2}$
(upper-left panel of Fig.~\ref{plot_en60}). But, already by $\mathbf{q}%
^2_{cm}=1$ fm$^{-2}$, the Bonn-potential and $\chi$EFT predictions (both
with FSI included) are within 10\% of each other, with the difference
entirely accounted for by the $\chi$EFT result's variation with the LSE
cutoff $\Lambda$. As we move to the quasi-free ridge, and FSI becomes less
important, this variation becomes less of a component of the full answer for 
$f_L$. In consequence there is a window, up to $\mathbf{q}^2_{cm} \approx 10$
fm$^{-2}$, where the cutoff dependence of the final prediction for $f_L$ is
not sizable. The agreement between $\chi$EFT and the Bonn potential for $f_L$
is also good through much of this range. However as $\mathbf{q}^2_{cm}$
approaches 10 fm$^{-2}$ the Bonn-potential and $\chi$EFT PWIA answers start
to differ, especially at small angles. Since the FSI is small in both
calculations in this range, that difference is not ameliorated in the full
calculation.

We will now summarize the results of the calculations we carried out for $%
E_{np}=10-160$ MeV and $\mathbf{q}_{cm}^{2}=0.1-25$ fm$^{-2}$. The results
are generically similar to those displayed above, in that, if we define%
\begin{equation}
\mathbf{q}_{qf}^{2}\text{ (fm}^{-2}\text{)}\equiv \frac{E_{np}\text{ (MeV)}}{%
10}
\end{equation}%
then our results show that for the kinematic region $E_{np}\le 60$ MeV and $|%
\mathbf{q}_{cm}^{2}-\mathbf{q}_{qf}^{2}|$ $<4$ fm$^{-2}$, the calculations
using $\chi$EFT up to NNLO have $<10$\% variation with respect to the cutoff
in the LSE . The $\chi$EFT results and those found with the Bonn potential
also agree within 10\% if $|\mathbf{q}_{cm}^{2}-\mathbf{q}_{qf}^{2}|$ $<2$ fm%
$^{-2}$, for $E_{np}$ at least as high as $160$ MeV. Somewhat remarkably
this agreement is possible even in cases where both $E_{np}$ and $\mathbf{q}%
^2_{cm}$ appear to be outside the range of validity of our calculation---as
long as we are close to the quasi-free ridge and so details of the FSI
remain unimportant.

On the other hand, we emphasize that the $\chi$EFT FSI is reliable for low $|%
\mathbf{q}_{cm}|$ and low $E_{np}$, see, e.g. first two panels of Fig.~\ref%
{plot_en10}. Indeed, in the first panel there one might be concerned about
the accuracy of our description of the deuteron wave function. The LECs in
our chiral potential are obtained through the best fit to the NN phase
shifts for $T_{lab}<100$ MeV. In general this does not give the best
possible deuteron wavefunction---which is crucial for the PWIA. An
alternative would be to perform the renormalization of the LECs in the $%
^{3}S_{1}-^{3}D_{1}$ channel so that some of the deuteron properties, e.g.,
the binding energy, are very accurately reproduced. We have verified that,
by doing this, for $|\mathbf{q}_{cm}^{2}-\mathbf{q}_{qf}^{2}|$ $<4$ fm$^{-2}$%
\ the 10\% uncertainty due to the variation of the cutoff in the LSE can be
reduced to 5\%, but only for lower values of $E_{np}(<30$ MeV$)$. Aside from
that, it does not improve the discrepancy between the results generated by $%
\chi$EFT and the Bonn potential. Moreover, for $E_{np}>60$ MeV, the
variation of the results with respect to the cutoff becomes larger than
before, due to the fact that the NN $^{3}S_{1}-^{3}D_{1}$ t-matrix now has
worse convergence with respect to the cutoff in the higher energy/momentum
region. We conclude that the uncertainty of $f_{L}$\ will remain roughly the
same unless higher orders in the chiral potential are included.

\begin{table}[tbph]
\begin{tabular}{|l|cccc|cc|}
\hline
& $p_{m}$ (MeV) & $|\mathbf{q}_{lab}|$ (MeV) & $E_{np}$ (MeV) & $|\mathbf{q}%
_{cm}|^{2}$ (fm$^{-2}$) & $f_{L}/f_{L}^{PWIA}$ & $f_{L}^{PWIA}$ (fm) \\ 
\hline
van der Schaar & 39.8 & 329 & 42.9 & 2.66 & 0.96 $\pm $ 0.03 & 5.00 \\ 
et al. & 69.9 & 299 & 49.5 & 2.18 & 0.93 $\pm $ 0.07 & 1.49 \\ 
& 110 & 259 & 59.3 & 1.62 & 0.77 $\pm $ 0.14 & 0.36 \\ 
& -38.8 & 419 & 29.4 & 4.36 & 1.08 $\pm $ 0.03 & 3.15 \\ 
& 39.8 & 381 & 53.7 & 3.51 & 1.07 $\pm $ 0.03 & 4.83 \\ 
& -58.8 & 503 & 36.9 & 6.24 & 0.96 $\pm $ 0.07 & 1.00 \\ 
Ducret et al. & -20 & 300 & 17.5 & 2.26 & 0.77 $\pm $ 0.02 $\pm $ 0.04 & 7.91
\\ 
& -20 & 400 & 51.1 & 2.19 & 0.88 $\pm $ 0.02 $\pm $ 0.05 & 1.32 \\ 
& -20 & 500 & 52.2 & 6.06 & 0.92 $\pm $ 0.02 $\pm $ 0.05 & 6.98 \\ 
& -20 & 600 & 75.0 & 8.53 & 0.92 $\pm $ 0.02 $\pm $ 0.06 & 5.44 \\ 
& -20 & 670 & 92.7 & 10.4 & 0.87 $\pm $ 0.03 $\pm $ 0.08 & 4.65 \\ 
& -100 & 400 & 10.3 & 4.06 & 0.88 $\pm $ 0.02 $\pm $ 0.06 & 0.137 \\ 
& -100 & 500 & 22.4 & 6.26 & 1.14 $\pm $ 0.02 $\pm $ 0.08 & 0.135 \\ 
& -100 & 600 & 38.5 & 8.86 & 1.15 $\pm $ 0.03 $\pm $ 0.10 & 0.114 \\ 
& -100 & 670 & 52.0 & 10.9 & 1.20 $\pm $ 0.03 $\pm $ 0.12 & 0.097 \\ 
& 100 & 200 & 41.8 & 0.981 & 0.51 $\pm $ 0.01 $\pm $ 0.03 & 0.493 \\ 
& 100 & 300 & 64.1 & 2.16 & 0.66 $\pm $ 0.01 $\pm $ 0.04 & 0.492 \\ 
& 100 & 400 & 90.1 & 3.73 & 0.75 $\pm $ 0.01 $\pm $ 0.04 & 0.444 \\ 
& 100 & 500 & 119 & 5.66 & 0.88 $\pm $ 0.02 $\pm $ 0.06 & 0.385 \\ 
Jordan et al. & 53 & 402 & 65 & 3.9 & $f_{L}=$ 1.78 $\pm $ 0.07 $\pm $ 0.15
& N/A \\ \hline
\end{tabular}%
\caption{The first two columns give kinematics quoted in the relevant
papers: the missing momentum, and the three-momentum transfer to the nucleus
as measured in the lab. frame. The next two columns give the (c.m. frame)
quantities we employ for our calculation. The fifth column shows results for 
$f_{L}/f_{L}^{PWIA}$ were extracted from plots given in Refs.~\protect\cite%
{vdS91,Du94,Jo96}. The first error is the statistical error, while the
second is the systematic quoted in the publication. The last column shows $%
f_{L}^{PWIA}$ evaluated with the Paris potential. The units for the Jordan
et al. $f_L$ measurement are fm.}
\label{tab-data}
\end{table}

\begin{table}[htbp]
\begin{tabular}{|l|cc|c|c|}
\hline
& $E_{np}$ (MeV) & $|\mathbf{q}_{cm}|^2$ (fm$^{-2}$) & $f_L$ (fm) & $%
f_L^{\chi EFT}$ (fm) \\ \hline
van der Schaar et al. & 42.9 & 2.66 & 4.59 $\pm$ 0.15 & 4.06 $\pm$ 0.04 \\ 
& 49.5 & 2.18 & 1.31 $\pm$ 0.10 & 1.09 $\pm$ 0.01 \\ 
& 59.1 & 1.62 & 0.265 $\pm$ 0.048 & 0.218 $\pm$ 0.005 \\ 
& 29.4 & 4.36 & 3.46 $\pm$ 0.10 & 2.78 $\pm$ 0.03 \\ 
& 53.7 & 3.51 & 4.84 $\pm$ 0.14 & 3.92 $\pm$ 0.03 \\ 
& 36.9 & 6.24 & 0.998 $\pm$ 0.073 & 0.911 $\pm$ 0.006 \\ 
Ducret et al. & 17.5 & 2.26 & 6.07 $\pm$ 0.24 & 6.52 $\pm$ 0.06 \\ 
& 51.1 & 2.19 & 1.09 $\pm$ 0.04 & 0.94 $\pm$ 0.01 \\ 
& 52.2 & 6.06 & 6.63 $\pm$ 0.22 & 6.36 $\pm$ 0.06 \\ 
& 75.0 & 8.53 & 5.50 $\pm$ 0.18 & 5.23 $\pm$ 0.05 \\ 
& 92.7 & 10.4 & 4.61 $\pm$ 0.16 & 5.02 $\pm$ 0.81 \\ 
& 10.2 & 4.06 & 0.119 $\pm$ 0.004 & 0.096 $\pm$ 0.003 \\ 
& 22.4 & 6.26 & 0.155 $\pm$ 0.005 & 0.116 $\pm$ 0.002 \\ 
& 38.5 & 8.86 & 0.136 $\pm$ 0.005 & 0.105 $\pm$ 0.001 \\ 
& 52.0 & 10.9 & 0.124 $\pm$ 0.004 & 0.091 $\pm$ 0.001 \\ 
& 41.7 & 0.981 & 0.245 $\pm$ 0.010 & 0.271 $\pm$ 0.005 \\ 
& 64.1 & 2.16 & 0.306 $\pm$ 0.015 & 0.326 $\pm$ 0.006 \\ 
& 90.1 & 3.73 & 0.298 $\pm$ 0.012 & 0.310 $\pm$ 0.004 \\ 
& 119 & 5.66 & 0.280 $\pm$ 0.013 & 0.261 $\pm$ 0.002 \\ 
Jordan et al. & 65.2 & 3.9 & 1.78 $\pm$ 0.07 & 2.15 $\pm$ 0.02 \\ \hline
\end{tabular}%
\caption{The experimental number quoted in the second-last column is the
result of multiplying the final two rows of the previous table (with the
exception of the Jordan et al. point). Only statistical errors are listed
for the experimental result presented in this table, since systematic errors
are accounted for separately. $f_L^{\protect\chi EFT}$ is an average of $%
\protect\chi$EFT calculations that include both PWIA and FSI pieces and use
cutoffs ranging from 600--1000 MeV, with the error bar indicating the size
of the spread.}
\label{tab-comparison}
\end{table}

Finally, we compare our results with data for the longitudinal structure
function published in Refs.~\cite{vdS91,Du94,Jo96}. The kinematics and data
for $f_L$ are listed in Table~\ref{tab-data}. The experiments of van der
Schaar et al.~\cite{vdS91} and Ducret et al.~\cite{Du94} give their data as
ratios between the measured longitudinal response and that predicted by a
Paris-potential~\cite{La80} impulse-approximation calculation. For those two
experiments we have calculated $f_L^{PWIA}$ at the pertinent momentum using
the Paris-potential deuteron wave-function parameterization of Ref.~\cite%
{La81}. The result for $f_L$ given in Table~\ref{tab-comparison} is then
obtained by multiplying the final two columns in Table~\ref{tab-data}
together (with the obvious exception of the single data point of Jordan et
al.~\cite{Jo96}, where the publication gives $f_L$ directly). Table~\ref%
{tab-comparison} compares these experimental results for $f_L$ with those
from our $\chi$EFT calculations using the SFR NNLO potential and cutoffs of
0.6 to 1 GeV.

In fact, we find significant sensitivity of both the Paris IA and the $\chi$%
EFT result to the value of $E_{np}$ chosen. The value of $E_{np}$ listed
here is obtained from the initial-state kinematics given in the
publications, using the relativistic energy-momentum relation. Using the
value of $E_{np}$ found from the final-state kinematics to compute $%
f_L^{PWIA}$ produces results that differ by more than the quoted
uncertainties on $f_L/f_L^{PWIA}$. This makes us think that further details
of the experiments, e.g., acceptances, are needed in order to provide a
completely meaningful comparison between theory and data.

Nevertheless, some trends are already clear from the straightforward
comparison between theory and data that can be made without detailed
knowledge of the experiments. For the $p_m=-20$ MeV data (Ducret et al.) the 
$\chi$EFT calculations are in good agreement with the data, once the
experiment's $\approx 6$\% systematic is accounted for. For example,
multiplying the Ducret et al. data by 1.015, we find that the $\chi$EFT
prediction is within 1.3 (combined theory and statistical uncertainty of the
measurement) $\sigma$ for all but the $E_{np}=51.1$ MeV data point. The
agreement for $p_m=100$ MeV is also very good---even at the highest $\mathbf{%
q}_{cm}^2$ of almost 6 fm$^{-2}$. For the rest of the Ducret et al. data
set, taken at $p_m=-100$ MeV, the $\chi$EFT calculation is systematically
below the data, and this cannot be attributed to normalization
uncertainty---at least not within the quoted systematic.

Similarly, the van der Schaar et al. data are (with one exception) all
underpredicted by $\chi$EFT at this order. This problem persists even if the
entire 7\% systematic uncertainty quoted in Ref.~\cite{vdS91} is assigned to
the experimental normalization. The disagreement worsens as $\mathbf{q}%
_{cm}^2$ increases, with the only experimental point for $\mathbf{q}_{cm}^2
< 2$ fm$^{-2}$ agreeing with the $\chi$EFT prediction while points in the
range $2~\mathrm{fm}^{-2} < \mathbf{q}_{cm}^2 < 3~\mathrm{fm}^{-2}$ have $%
\chi$EFT 1-3$\sigma$ below the data (depending on how the experimental
normalization is treated). The disagreement between $\chi$EFT and data
becomes dramatic at higher $\mathbf{q}_{cm}^2$.

Similar under-prediction also occurs when Arenh\"ovel's Bonn-potential
calculation is compared to these data, so the difficulty appears generic to
calculations employing only the one-body operator for $J_0$. It will be
interesting to explore whether higher-order corrections to the NN charge
operator in $\chi$EFT can redress the difference seen here between theory
and experiment. It is also worth noting that the uncertainty due to cutoff
variation in the $\chi$EFT result is very small for most cases. This leads
us to question whether varying the cutoff between 600 MeV and 1 GeV
adequately estimates the uncertainty in the $\chi$EFT result due to
higher-order effects.

\section{Conclusion}

\label{sec-conc}

We have computed the longitudinal structure function of the deuteron, $f_{L}$%
, in $\chi $EFT, including effects up to $\mathcal{O}(P^{3})$ relative to
leading order in the standard counting.
Comparison with calculations of Arenh\"{o}vel \textit{et al.}, which use the
Bonn potential, indicates good agreement at low energies ($E_{np}\leq 60$
MeV) and for $\mathbf{q}^{2}$ which are not very distant ($<4$ fm$^{-2}$)
from the quasi-free ridge. We also use the $\Lambda $ dependence of the $%
\chi $EFT calculation as a diagnostic, to find kinematic regions where the
final result for $f_{L}$ is sensitive to short-distance components in the
evaluation of the final-state interaction. If significant $\Lambda $
dependence is present it suggests sensitivity to such effects, which may
mean that both the $\chi $EFT calculation and the Bonn-potential calculation
do not capture the full dynamics present in $f_{L} $. With this caveat in
mind, we find that our calculations are able to describe data from Saclay,
NIKHEF, and Bates, on the longitudinal structure function, either within a
more restricted kinematic region (compared to the region where our results
agree with Bonn potential), or by allowing somewhat expanded combined
(statistical + systematic + theory) error bars. We also notice that both $%
\chi $EFT and Bonn potential give a similar trend of under-prediction of the
experimental $f_{L}$ with increasing $\mathbf{q}_{cm}^{2}$. Further studies
are needed to understand the origin of this discrepancy.

In fact, the $\chi$EFT calculation is in good agreement with the Bonn
potential and with the data to surprisingly high energies and $\mathbf{q}^2$
as long as one stays on (or near) the quasi-free ridge. This suggests that a
different counting may be needed for this reaction, one where the expansion
is not in the naive kinematic variables, $|\mathbf{p}^{\prime}|$ and $|%
\mathbf{q}|$, but instead, perhaps where some account is taken of the
dominance of the PWIA part of the matrix element for sufficiently high
energies ($E_{np} > 30$ MeV) and sufficiently low values of the ``missing
momentum" $|\mathbf{p} - \frac{\mathbf{q}}{2}|$.

In the future, the recent $\chi $EFT calculations of elastic scattering on
tri-nucleons~\cite{Pa12} could be extended to electrodisintegration, and
compared with data on that reaction~\cite{exp_dis_3n}. However, a more
obvious and immediately necessary next step is to extend this $\chi $EFT
calculation to other structure functions. The three-current operator $%
\mathbf{J}$ has also been computed to three orders relative to leading order~%
\cite{Ko09,Ko11,Pa09,Pa11}, and comparison of $f_{T}$, $f_{LT}$, and $f_{TT}$
with data (and Bonn-potential calculations) would be an important test of
the domain of validity of the $\chi $EFT expansion for that object.

In closing, we reiterate that such a test can be rendered unambiguous
because the results presented here show the regions in which the $\chi$EFT
expansion for the PWIA and FSI pieces of the deuteron electrodisintegration
process are under control. The absence of any NN mechanisms in the charge
operator up to the order considered makes the $f_L$ presented here a
prediction once the NN potential is fixed by the fit to NN data. The fact
that we find good agreement with both $f_L$ data and other theories in a
broad kinematic range provides further reassurance (see also Refs.~\cite%
{Ph07,Ko12}) that $\chi$EFT does a good job of describing deuteron structure
for internal relative momenta $< 0.2$ GeV.


\acknowledgments We thank H. Arenh\"ovel for providing his Bonn potential
results for comparison. C.-J. Yang thanks B. Barrett, C. Elster, S. Fleming
and U. van Kolck for vaulable support. This work is supported by the US NSF
under grant PHYS-0854912 and US DOE under contract No. DE-FG02-04ER41338 and
DE-FG02-93ER40756.

\end{document}